\documentstyle[psfig]{mn2e}

\def\gs{\mathrel{\raise0.35ex\hbox{$\scriptstyle >$}\kern-0.6em
\lower0.40ex\hbox{{$\scriptstyle \sim$}}}}
\def\ls{\mathrel{\raise0.35ex\hbox{$\scriptstyle <$}\kern-0.6em
\lower0.40ex\hbox{{$\scriptstyle \sim$}}}}
\def\ls{\mathrel{\hbox{\rlap{\hbox{\lower4pt\hbox{$\sim$}}}\hbox{$<$}}}}
\def\gs{\mathrel{\hbox{\rlap{\hbox{\lower4pt\hbox{$\sim$}}}\hbox{$>$}}}}

\title[Stellar mass and environment in Abell~1691]
{The role of stellar mass and environment for cluster blue fraction,
AGN fraction and star-formation indicators from a targeted analysis of Abell~1691}
\author[Pimbblet and Jensen]
       {Kevin A.\ Pimbblet$^{1,2}$\thanks{email: Kevin.Pimbblet@monash.edu} 
and Peter C.\ Jensen$^{3}$
        \vspace*{1mm}\\
        $^{1}$School of Physics, Monash University, Clayton, Victoria 3800, Australia\\
$^{2}$Monash Centre for Astrophysics (MoCA), Monash University, Clayton, Victoria 3800, Australia\\
$^{3}$Centre for Astrophysics and Supercomputing, Swinburne University of Technology,
Hawthorn, Victoria 3122, Australia
}

\date{\today}

\pagerange{000--000}

\begin{document}

\maketitle

\begin{abstract}
We present an analysis of the galaxy population of the 
intermediate X-ray luminosity galaxy cluster, Abell~1691, 
from SDSS and Galaxy Zoo data to elucidate the relationships
between environment and galaxy stellar mass for a variety of 
observationally important cluster populations that include
the Butcher-Oemler blue fraction, the active galactic nucleus (AGN) 
fraction and other spectroscopic classifications of galaxies.  
From 342 cluster members, we determine a cluster
recession velocity of $21257\pm54$ kms$^{-1}$ and velocity
dispersion of $1009^{+40}_{-36}$ kms$^{-1}$ and show that
although the cluster is fed by multiple filaments of galaxies
it does not possess significant sub-structure in its core.
We identify the AGN population of the cluster from a BPT 
diagram and show that there is a mild increase in the AGN 
fraction with radius from the cluster centre that appears
mainly driven by high mass galaxies (log(stellar mass)$>10.8$).  
Although the cluster blue fraction follows the same radial 
trend, it is caused primarily by lower mass
galaxies (log(stellar mass)$<10.8$).
Significantly, the galaxies that have undergone recent 
star-bursts or are presently star-bursting but dust-shrouded
(spectroscopic e(a) class galaxies) are also
nearly exclusively driven by low mass galaxies.  We therefore
suggest that the Butcher-Oemler effect may be a
mass-dependant effect.  
We also examine red and passive spiral galaxies and show that
the majority are massive galaxies, much like the rest of the
red and spectroscopically passive cluster population.  
We further demonstrate that
the velocity dispersion profiles of low and high mass
cluster galaxies are different.  Taken together, we infer that the duty 
cycle of high and low mass cluster galaxies are markedly different, with
a significant departure in star formation and specific star formation
rates observed beyond $r_{200}$
and we discuss these findings.

\end{abstract}


\section{Introduction}
Arguably the most important, emergent conclusion about galaxy evolution
during the past decade is that the two main driving forces behind 
it are the stellar mass of a galaxy and its large-scale environment. 
Although stellar mass may be the more important of the pair,
the environmental effects are also non-negligible and the two variables
likely exhibit a significant covariance 
(Baldry et al.\ 2006; see also von der Linden et al.\ 2010).

\begin{figure*}
\vspace*{-8cm}
\centerline{\psfig{file=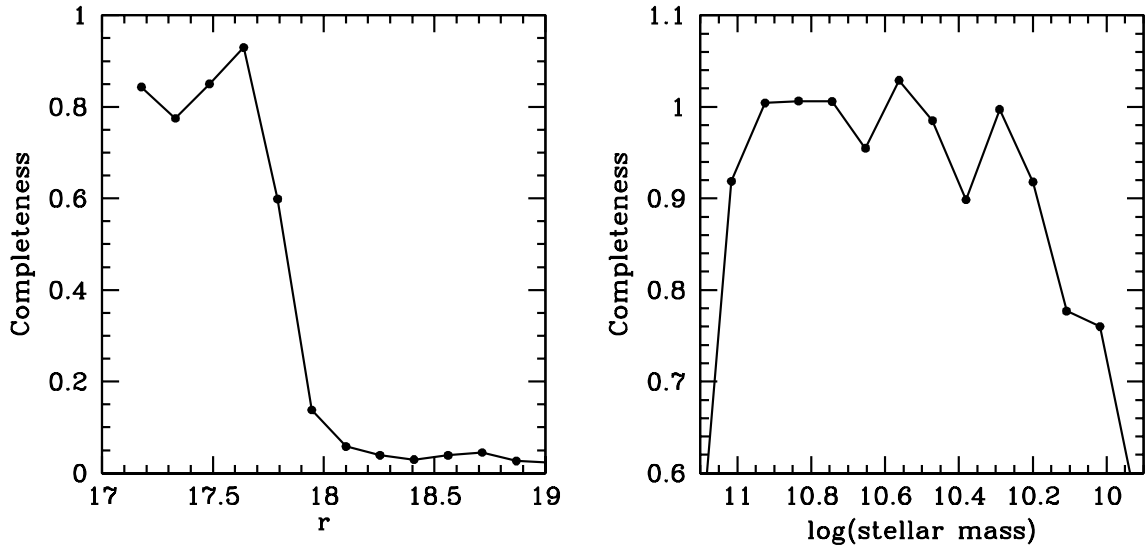,angle=0,width=6.5in}}
\caption{Completeness for our galaxy sample in terms of 
apparent r-band magnitude (left) and stellar mass (right). 
At $r=17.7$ and log(stellar mass)=10.2, the data is 
$\sim90$ per cent complete.
}
        \label{fig:compl}
\end{figure*}

For example, in the extreme (and hostile) environment of
the dense cores of massive galaxy clusters
galaxy colour is preferentially red 
and galaxy morphology elliptical -- a systematic
difference to non-cluster galaxies 
(e.g.\ Dressler 1980;
Dressler et al.\ 1997;
Lewis et al.\ 2002;
G{\'o}mez et al.\ 2003;
Balogh et al.\ 2004a; 2004b;
Kauffmann et al.\ 2004;
Christlein \& Zabludoff 2005;
Baldry et al.\ 2006;
Cooper et al.\ 2006
Sato \& Martin 2006;
Pimbblet et al.\ 2006;
Bamford et al.\ 2009;
Skibba et al.\ 2009;
Deng et al.\ 2010;
Peng et al.\ 2010;
Ma \& Ebeling 2011;
Patel et al.\ 2011;
Calvi et al.\ 2012;
Jensen \& Pimbblet 2012).

The mass function of cluster galaxies 
shows that the more massive galaxies are also redder
and more frequently elliptical
(e.g.\ 
Baldry et al.\ 2006;
di Serego et al.\ 2006;
Clemens et al.\ 2006;
van der Wel 2007;
van den Bosch et al.\ 2008;
Bamford et al.\ 2009;
Deng et al.\ 2010;
Pasquali et al.\ 2010;
Peng et al.\ 2010;
Nair \& Abraham 2010;
Thomas et al.\ 2010;
Gr{\"u}tzbauch et al.\ 2011;
Giodini et al.\ 2012;
Wilman \& Erwin 2012;
see also Vulcani et al.\ 2012).

Early investigations in to galaxy evolution inside
clusters of galaxies by Butcher \& Oemler (1978; 1984)
suggested that the fraction
of blue galaxies decreases rapidly with decreasing redshift.
They hypothesized that the cause of this dramatic change in blue
fraction to be due to spiral galaxies undergoing ram-pressure stripping
caused by interaction with the intra-cluster environment (e.g.\ Gunn \& Gott 1972).  
Such sprials
morphologically transform in to lenticular (S0) galaxies and
fade as they are no longer able to replenish their gas supplies.
Later works confirmed the existence of the Butcher-Oemler effect
(Couch et al.\ 1994;
Rakos \& Schombert 1995; 
Lubin 1996;
Dressler et al.\ 1997;
Smail et al.\ 1998;
Ellingson et al.\ 2001; 
Kodama \& Bower 2001; 
Margoniner et al.\ 2001;
Goto et al.\ 2003;
Tran et al.\ 2003;
De Propris et al.\ 2004;
Urquhart et al.\ 2010).
Yet observational follow-up by Dressler \& Gunn (1983) demonstrated
that the blue galaxies measured by the Butcher-Oemler effect are
actually undergoing star-bursts rather than gently fading away, indicating
that instead of gas being stripped in a galaxy, in could be used up
in a spectacular, albeit terminal, starburst phase.
Couch \& Sharples (1987) made an independent spectroscopic study 
of $z\approx0.3$ clusters and came to the same conclusion: the 
blue galaxies are being driven by galaxies that have undergone
a recent (explicitly: within 0.5 Gyr) starburst.  
This observational conclusion has been re-enforced by
later studies (e.g.\ Lavery \& Henry 1988; 
Fabricant et al.\ 1994;
Couch et al.\ 1994; 1998;
Abraham et al.\ 1996;
Fisher et al.\ 1998;
Poggianti et al.\ 1999;
Dressler et al.\ 2004).
Moreover,
some of the other cluster galaxies also exhibited the rare 
signatures of a post-starburst phase 
(E+A galaxies; e.g.\ Zabludoff et al.\ 1996) suggesting a duty cycle
for these blue galaxies (see also Mahajan et al.\ 2011).
The cause of the blue Butcher-Oemler galaxies has subsequently 
been argued to be due to a number of different physical mechanisms 
including, but not limited to, 
galaxy-galaxy interaction 
(Lavery \& Henry 1988; Couch et al.\ 1994;
Moore et al.\ 1996; Quilis et al.\ 2000);
galaxy-cluster interaction (Henriksen \& Byrd 1996; 
Bekki et al.\ 2001);
galaxy mergers (van Dokkum et al.\ 1999; Dressler et al.\ 1999);
or a slow starvation of star-forming gas 
(Larson, Tinsley \& Caldwell 1980; 
Balogh \& Morris 2000;
Bekki, Couch \& Shioya 2002). 
These mechanisms operate most efficiently in different environs
which would lead to differentiable observational signatures (e.g.\
ram pressure induced starbursts should cause more blue galaxies close
to the cluster core, whereas mergers would not do so).

\begin{figure*}
\vspace*{-3cm}
\centerline{\psfig{file=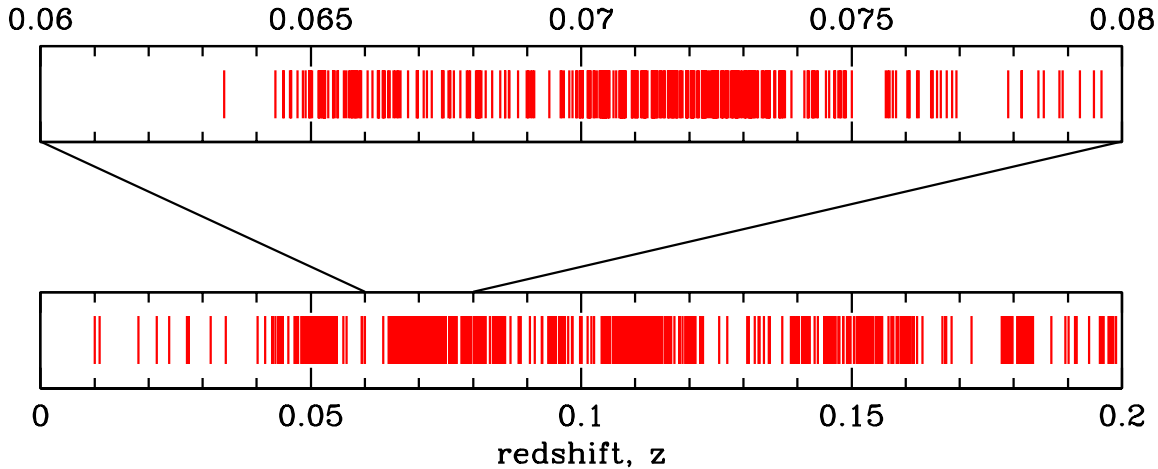,angle=0,width=6.75in}}
\vspace*{-7cm}
\caption{Barcode-style redshift diagram for all galaxies within 2 degrees
of Abell 1691.  Each vertical line represents an individual galaxy.  In the
upper plot, we show a zoom-in on the $0.06<z<0.08$ region of the lower plot.
We use this plot to estimate the upper and lower bounds of the cluster
in order to derive its mean redshift and velocity dispersion.
}
        \label{fig:barcode}
\end{figure*}

We caution here, however, that there are certainly observational biases affecting
some studies (as indicated by Andreon \& Ettori 1999; 
De Propris et al.\ 2003; Andreon et al.\ 2006; Holden et al.\ 2007),
claims of a lack of Butcher-Oemler effect (e.g.\ Haines et al.\ 2009;
Andreon et al.\ 2006), or at the very least a dampened effect when some
of these selection biases are taken in to account 
(cf.\ Margoniner \& de Carvalho 2000).  For example, if galaxies are
mass-selected, then bluer galaxies are seldom
found in more massive regimes ($>10^{11}$ solar masses; see Raichoor \&
Andreon 2012a,b).

Taking the above results together, it is suggestive that a starbursting
phase is very important in the evolution (or duty-cycle) 
of cluster galaxies (Oemler et al.\ 2009) -- especially at lower masses 
(Mahajan et al.\ 2011).  
As galaxies accrete on to the cluster potential, \emph{some} may experience
a starburst due to one or more of the above mechanisms 
(see also Porter et al.\ 2008; Fadda et al.\ 2008).
Whilst galaxies outside a cluster tend to
revert to their long-term star-formation
rates after a starburst, cluster galaxies do not (cf.\ Poggianti et al.\ 1999;
Dressler et al.\ 2004; Oemler et al.\ 2009; see also Dressler et al.\ 2009).
The duty cycle of cluster galaxies therefore merits further investigation 
at all redshifts
-- particularly if we are to elucidate the connections between mass,
environment, colour and morphology that we have summarized above.
We can also tie these relationships to incidence of active galactic
nuclei (AGN).
From a simplistic observational viewpoint, powerful AGN are found
in massive galaxies with current, on-going star-formation 
(e.g.\ Magorrian et al.\ 1998;
Kauffmann et al.\ 2003; Heckman et al.\ 2005;
von der Linden et al.\ 2010; Xue et al.\ 2010; see also Aird et al.\ 2012).  
Since AGN require a fuel source (i.e.\ gas), any physical mechanism that 
disturbs the morphology of (spiral) galaxies may
also directly enhance AGN activity.  
Alexander \& Hickox (2012; and references therein) give a review of how 
the AGN central engines can grow, such as via direct galaxy mergers.
For a mechanism such as
galaxy-galaxy harassment (Moore et al.\ 2006), 
AGN may trace galaxy interactions and constitute a smoking-gun for
galaxy evolution (see also Lake et al.\ 1998).  The establishment of a link between
AGN incidence and environment is still in the balance, however.
Whilst some researchers contend that AGN incidence is affected
by environment 
(Kauffmann et al.\ 2004;
Ruderman \& Ebeling 2005;
Popesso \& Biviano 2006;
Gilmour et al.\ 2007;
Constantin et al.\ 2008; 
Sivakoff et al.\ 2008;
Montero-Dorta et al.\ 2009;
Gavazzi et al.\ 2011)
and there may be a `Butcher-Oemler AGN effect' 
(Eastman et al.\ 2007; Aird et al.\ 2012),
others come to the opposite conclusion
(Miller et al.\ 2003;
Martini et al.\ 2007;
Georgakakis et al.\ 2008;
Sivakoff et al.\ 2008;
Pasquali et al.\ 2009;
Haggard et al.\ 2010;
von der Linden et al.\ 2010;
Atlee et al.\ 2011).

It is under these considerations that we herein  
present an analysis of the galaxy population of
Abell~1691 based on Sloan Digital
Sky Survey data (Abazajian et al.\ 2009).
Abell~1691 is an intermediate X-ray luminosity cluster 
($L_X = 0.889 \times 10^{44}$ ergs$^{-1}$; Ebeling et al.\ 2000;
see Jensen \& Pimbblet 2012 for an introduction to our intermediate
$L_X$ studies)
and as such, may not constitute as extreme an environment as 
higher X-ray luminosity clusters that have been studied in 
detail and extensively elsewhere (Pimbblet et al.\ 2006).
For example, Jensen \& Pimbblet (2012) find that the modal colour
of red sequence galaxies in intermediate X-ray luminosity
clusters at $z\sim0.08$ is less than for higher X-ray
luminosity (i.e.\ higher mass) galaxy clusters (Pimbblet et al.\ 2006).
The reason for selecting Abell~1691 for a single-case study is that
it contains a large number of spectroscopically confirmed redshifts
within $3r_{virial}$ (see Section~3 below, and Jensen \& Pimbblet 2012)
and is a relatively under-studied cluster in the literature but
representative of the intermediate $L_X$ ranges investigated 
in our previous work (Jensen \& Pimbblet 2012).

The datasets that we use in this work are described in Section~2.
We begin our analysis of Abell~1691 in Section~3 by determining 
cluster membership from a gapping approach and searching for any
large-scale sub-structuring.  In Section~4, we examine how the
cluster environment affects active galactic nuclei cluster members.
We complement this with an analysis of the colour-magnitude relation
of the cluster in Section~5.  In Section~6 we apply a spectroscopic
classification scheme to the cluster members to examine the duty cycle
of the cluster's galaxies.  This is complemented with an investigation
of the environment of red and passive spiral galaxies.  In
Section~7 we determine how mass drives AGN fraction, blue fraction
and spectroscopic duty cycle before summarizing our main
conclusions in Section~8.

Throughout this work, we adopt 
a standard, flat cosmology with 
$\Omega_M = 0.238$, $\Omega_{\Lambda}=0.762$ and $H_0=73$ km s$^{-1}$ Mpc$^{-1}$
(Spergel et al.\ 2007).

\section{Dataset}	

In this work, we make use of SDSS Data Release 7 (Abazajian et al.\ 2009).
We download all galaxies that have had spectroscopic observations taken
within 2 degrees of the nominal cluster centre of Abell 1691 (as specified
by the NASA Extra-galactic Database; NED).
We note that the SDSS spectroscopy 
aims to be 100 per cent complete down to a magnitude
limit of $r=17.77$ for all sky areas (Strauss et al.\ 2002).  This magnitude limit
corresponds to $\sim M^{\star} + 1$ at the redshift of the cluster.
In reality, the $r=17.77$ target selection
may not be 100 per cent spectroscopically complete due to 
the limitations of how close pairs of spectra can be 
observed (Blanton et al.\ 2003).  The interlacing of latter spectroscopic
observations may recover some of these collision limitations, but the final
completeness requires investigation (Fig.~\ref{fig:compl}).

At various points in the subsequent analysis we restrict 
ourselves to $r=17.77$ to maximize the number of redshifts we work with
and probe down to a mass of log(stellar mass)$\approx10$ 
(where stellar
mass has units of solar masses throughout this work). 
We evaluate how complete we are at these magnitudes by determining the
fraction of spectroscopic targets that are observed by SDSS and generate a
successful redshift compared to how many are potentially available for
spectroscopic observation in Fig.~\ref{fig:compl}.
We find that at $r=17.77$, we are $\sim90$ per cent complete.  
To evaluate how complete we are in stellar mass for this sample, we
construct histograms of stellar mass and fit a log-linear relation to
the range 11$<$log(stellar mass)$<$10.5 (where we anticipate being
$\sim100$ per cent complete) and extrapolate to lower masses.
We find that our catalogue is $>90$ per cent complete at 
log(stellar mass)$= 10.2$.
We contend that these limits in apparent $r$ band magnitude and
stellar mass are more than satisfactory for the purposes of the present work.

In addition, we marry the dataset up with line measurements
derived by Tremonti et al.\ (2004) to provide equivalent width
values for important spectral lines that we use in this work
and star formation rates from Brinchmann et al.\ (2004).
This is supplemented by morphological data from the 
Galaxy Zoo project (Lintott et al.\ 2008; 2011).
In brief, Galaxy Zoo is an online `citizen science' undertaking
that invites members of the public to morphologically classify
SDSS galaxies in to several categories. The classifications agree
with professional astronomers to an accuracy of better than 90 per cent
(Lintott et al.\ 2008).  That said, we caution that the resolution of 
Galaxy Zoo may cause a blending of some classes (particularly 
at the Sa/S0 borderline; cf.\ Andreon \& Davoust 1997).  Given
these caveats and the fact 
that the Galaxy Zoo morphologies compare well with
various external datasets (Lintott et al.\ 2008), 
we limit ourselves in this work to dividing our galaxies in to
three categories: spiral, elliptical and uncertain from Galaxy Zoo.

\section{Cluster Membership}

Cluster membership for Abell 1691 is determined from the SDSS
spectroscopy (Fig.~\ref{fig:barcode}).  There are a number
of techniques for defining cluster membership from such data,
but one of the most widely used and simplest is a sigma-clipping 
(or `gapping') approach
(Yahil \& Vidal 1977; Zabludoff, Huchra \& Geller 1990; see also
Pimbblet et al.\ 2006).  This approach takes the redshift distribution
and applies a sensibly-chosen upper and lower bound to the region
of interest (upper panel of Fig.~\ref{fig:barcode}).  The mean and 
standard deviation of these galaxies is then computed.  Any galaxy
found to be beyond $3\sigma_z$ of the mean is then eliminated from
the sample.  The process is then iterated.  

\begin{figure}
\centerline{\psfig{file=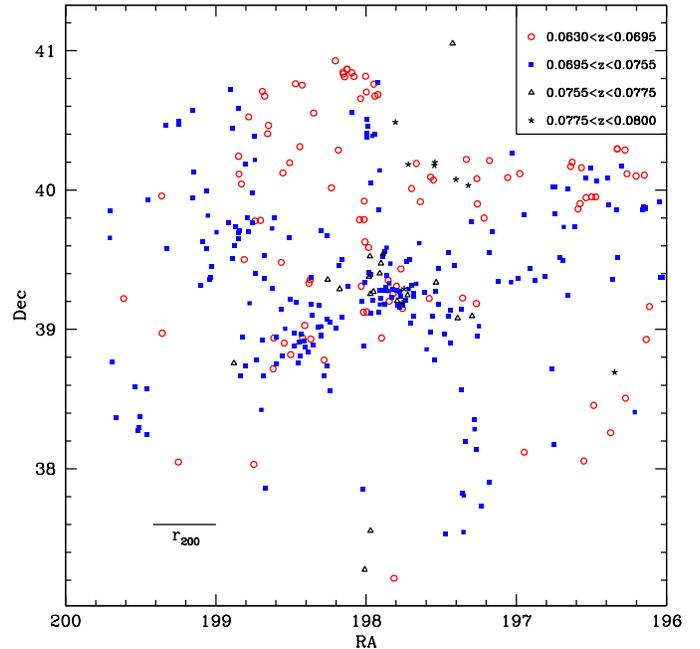,angle=0,width=3.75in}}
\caption{Spatial distribution of all galaxies 
in our sample with $0.063<z<0.080$, coded according to their redshifts.
We use this information to derive values for the cluster's
velocity dispersion in conjunction with the redshift distribution and
knowledge of $L_X$.  A scale bar showing $r_{200}$ ($=2.01$ Mpc) at
the redshift of the cluster is depicted in the lower left corner.
Several filaments can be seen extending from the centre of
Abell~1691 and are presumably fuelling the growth of this
cluster (cf.\ Pimbblet et al.\ 2004).  
}
        \label{fig:structure}  
\end{figure}

\begin{figure}
\centerline{\psfig{file=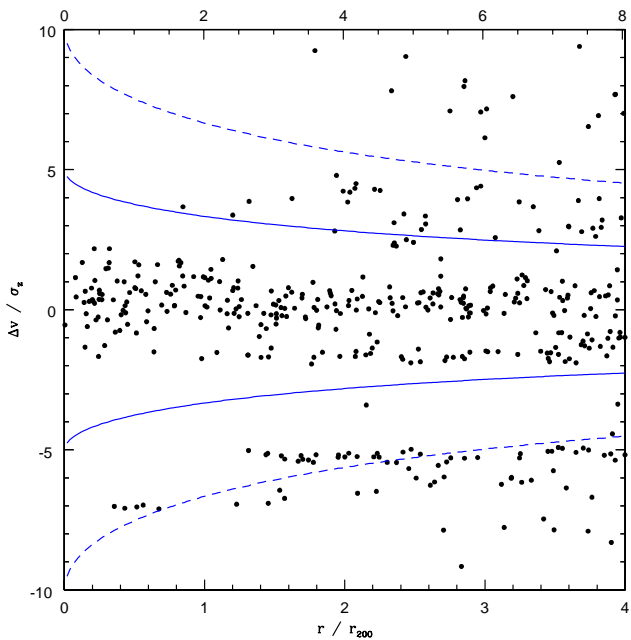,angle=0,width=3.5in}}
\caption{Phase-space diagram of velocity offset as a function
of distance from the cluster centre for the galaxies in our
sample.  The solid lines denote 3$\sigma$ caustic from the
mass model of Carlberg et al.\ (1997) and the dashed lines are
the 6$\sigma$ caustic.  Galaxies inside the solid line
are considered cluster members whilst those between the solid and
dashed lines can be thought of as infalling.  
}
        \label{fig:cye}
\end{figure}

\begin{figure}
\centerline{\psfig{file=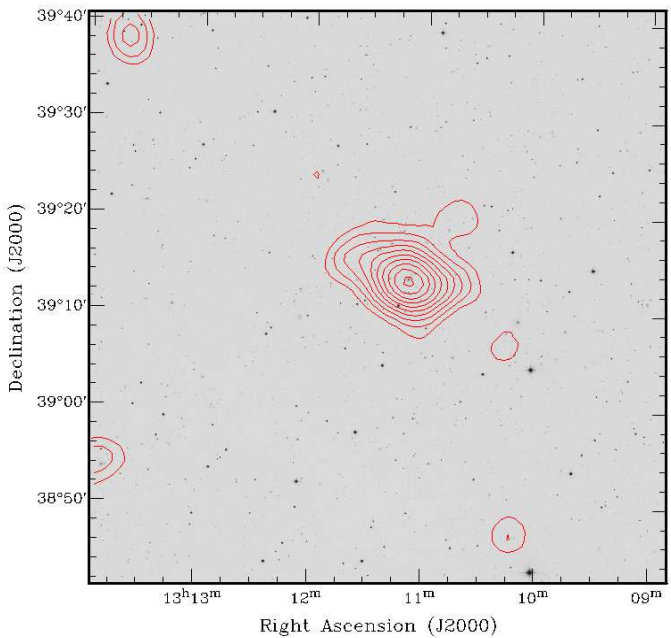,angle=0,width=3.5in}}
\caption{X-ray contour map generated from ROSAT 0.1--2.4 keV  
imaging overlaid on to the r-band image of Abell~1691.   This
suggests a cluster core region without significant substructure.  
}
        \label{fig:xray}
\end{figure}

At $z=0.06$ and $z=0.08$ there appear gaps in the redshift distribution
which we use as our first estimate of the upper and lower bounds for the
cluster.  Applying the clipping technique, we find that 
$\overline{cz}=21257 \pm 54$ kms$^{-1}$ with 
$\sigma_z = 1009^{+40}_{-36}$ kms$^{-1}$
where the errors are obtained from the method of
Danese, De Zotti \& di Tullio (1980).

However, a closer examination of the redshift distribution 
(Fig.~\ref{fig:barcode}) suggests that there may be a foreground
structure at $z\approx0.066$ which may not be part of the cluster proper.
Consequentially, if we choose a new lower bound of $z=0.0695$,
we obtain $\overline{cz}=21614 \pm 27$ kms$^{-1}$ with 
$\sigma_z = 402^{+21}_{-18}$ kms$^{-1}$.  These two approaches
clearly result in significantly different values of mean cluster redshift
and velocity dispersion.  To resolve which value we should adopt, 
we consider two factors: firstly, whether the foreground structure
is coherent (e.g.\ an infalling group or sub-cluster) and secondly,
an examination of cluster scaling relationships (Popesso et al.\ 2005).

For the first consideration, we create a map of RA versus Dec 
(Fig.~\ref{fig:structure}) and 
examine it for any structure in redshift bands suggested
to us by Fig.~\ref{fig:barcode}.
This suggests that the foreground redshift peak is smeared over the
sky and does not represent a group or cluster in its own right
(in contrast with Abell 1691 itself which is revealed as a
strong over-density in the centre of this plot, with a significant
secondary peak Eastward of the centre and filaments radiating outward).

For the second consideration, we note that the X-ray
luminosity of A1691 is $L_X = 0.889 \times 10^{44}$ ergs$^{-1}$ in 
the ROSAT 0.1--2.4 keV band (Ebeling et al.\ 2000).
Popesso et al.\ (2005) notes that 
$log_{10}(L_X) = 3.68 log_{10}(\sigma_z) -10.53$
for all galaxies in a given cluster (i.e.\ not just red early-types).
When we plug in our two values for $\sigma_z$
in to the above equation, it yields $L_X=3.3 \times 10^{44}$ ergs$^{-1}$
for $\sigma_z=1009$ kms$^{-1}$, and $L_X=0.1 \times 10^{44}$ ergs$^{-1}$
for $\sigma_z=402$ kms$^{-1}$.  Interestingly, the 
measured value for $L_X$ lies between these two values.
To resolve this, we note that
Popesso et al.\ (2005) quotes the 95\% confidence level on
the scatter of the intercept of this relationship as 0.80.
This generates a 95\% confidence interval
of $0.02 \times 10^{44} <L_X<0.7 \times 10^{44}$ ergs$^{-1}$ for
$\sigma_z=402$ kms$^{-1}$ and
$0.5 \times 10^{44} <L_X< 20.9 \times 10^{44}$ ergs$^{-1}$
for $\sigma_z=1009$ kms$^{-1}$.
This, coupled with Fig.~\ref{fig:structure}, suggests to us that 
the better redshift interval to derive the cluster parameters
from is $0.06<z<0.08$ and therefore we adopt
$\overline{cz} = 21257 \pm 54$ kms$^{-1}$ and
$\sigma_z = 1009^{+40}_{-36}$ kms$^{-1}$ for Abell 1691.

To check these findings, we finally construct a phase-space
diagram (Fig.~\ref{fig:cye})
of velocity versus distance from the centre of the cluster
for all galaxies in our magnitude-limited sample 
using the cluster mass model of Carlberg, Yee and Ellingson (1997).
In this, we see that the foreground structure noted above ($0.06<z<0.0695$)
mixes with the higher redshift range ($0.0695<z<0.0755$) at most radii
from the cluster centre and, moreover, 
is well contained within the $3\sigma$ caustic
contour from the Carlberg et al.\ (1997) mass model.  
We therefore adopt
our final definition of cluster members as being those galaxies contained
within these $3\sigma$ limits.  This yields a total of 342 galaxies 
for our study.

\subsection{Sub-structure}
\begin{figure*}
\centerline{\psfig{file=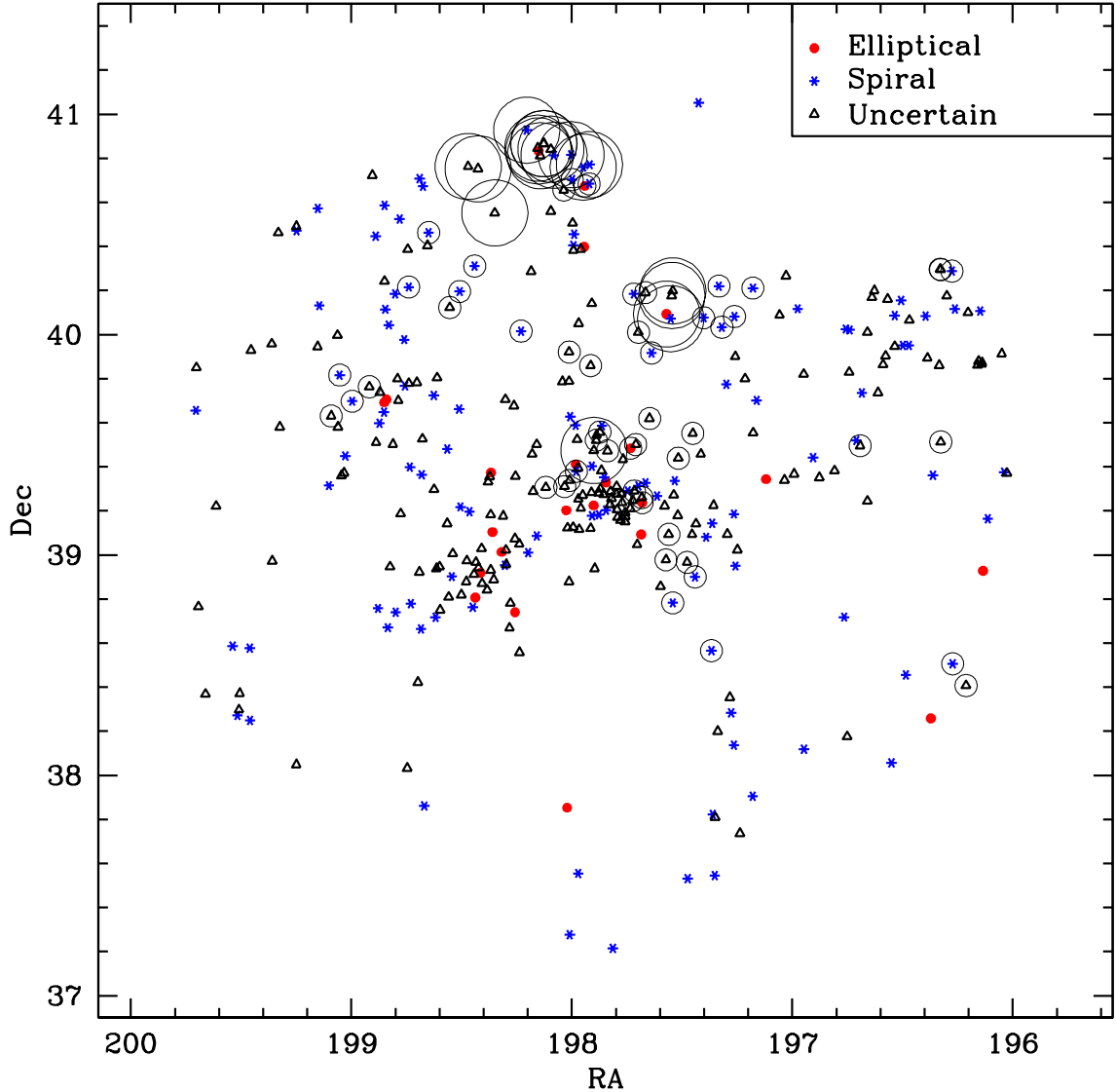,angle=0,width=6.5in}}
\vspace*{-1cm}
\caption{DS test result with galaxy members coded according
to their Galaxy Zoo morphologies (ellipticals are red circles; 
spirals are blue stars; uncertain morphologies are open black 
triangles).  Normally, one would plot circles around each point
that have a radius proportional to $e^{\delta}$.  However,
to improve clarity, we only plot two circle radii to denote modest
and significant values of $e^{\delta}$ (smaller and larger circles
respectively). 
Sub-structure is interpreted as overlapping circles in close proximity.
The main clump of sub-structure is seen to be north 
of the central overdensity (RA$\approx 198$, Dec$\approx$40.8)
and may be an infalling group.
}
        \label{fig:subs}
\end{figure*}

Although the cluster appears moderately smooth (with 
a moderate secondary 
overdensity eastward of the centre; Fig.~\ref{fig:structure}),
we aim to ascertain if there is bona-fide sub-structure 
contained within it.  We firstly turn to ROSAT X-ray imaging
of the galaxy cluster (Fig~\ref{fig:xray}) to check if the
cluster is experiencing any overt signs of recent interactions 
(cf.\ Owers et al.\ 2009; Oemler et al.\ 2009). The lack of
substructure, coupled with regular (i.e.\ roughly circular),
contours indicates that Abell~1691 is a virialized system that
hasn't had any recent perturbation from (e.g.) an infalling
system of comparable mass. 
But would we be able to detect sub-structure in this X-ray image?
Firstly, we note that spatial resolution of ROSAT PSPC is $\sim 25''$ at 0.93 keV
which translates to $\sim 33$ kpc at the redshift
of A1691. This is approximately the diameter of a single galaxy, hence we
believe that ROSAT has sufficient spatial resolution to resolve 
sub-structures at the redshift of A1691.
Secondly, although there is scatter in the $L_X$-mass relationship 
(Popesso et al.\ 2005), 
we note from Fig.~1 of Jensen \& Pimbblet (2012) that 
we would be able to detect clusters down to $L_X\sim0.7 \times 10^{-44}$ ergs$^{-1}$
from ROSAT. We explicitly note that that does not 
preclude a cluster of equal mass being present in our sample, 
but we regard it as unlikely.

To address this question of substructure further, we 
apply the Dressler \& Shectman (1988; DS) 
sub-structure test to our dataset.  This test is 
the most sensitive available to detect sub-structure
in the literature (Pinkney et al.\ 1996).  It works
by calculating the mean local velocity ($\overline{cz}$)
and local velocity standard deviation ($\sigma_{local}$)
of a given galaxies $N$ nearest neighbours (where $N$ is
chosen to be the 10 nearest neighbours).
To determine if there exists significant sub-clustering, 
these values are compared to the parent cluster's 
mean velocity and $\sigma_z$ such that:
\begin{equation}
\delta^2 = \left( \frac{ N_{local} + 1 }{\sigma_v^2} \right)
[ (\overline{cz}_{local}-\overline{cz})^2 + ( \sigma_{local}-\sigma_v)^2 ]
\end{equation}
where $\delta$ is a measure of the deviation of the 
individual galaxy.  The parameter of merit, $\Delta$, is
then the summation of all $\delta$ terms in the cluster.
This is compared to a Monte Carlo re-simulation of the
cluster wherein the velocities are randomly re-assigned
to each galaxy to generate $P(\Delta)$ and thereby 
estimate the confidence level that the cluster contains
sub-structure.  For Abell~1691, we find that 
$P(\Delta)<0.01$ suggesting it has significant
sub-structuring. This can be seen in Fig.~\ref{fig:subs}
where we plot circles 
on the basis of a galaxies' $e^{\delta}$
value -- overlapping large circles in close
proximity are therefore logically 
interpreted as sub-structure (see DS).

Although the cluster core ($r<\sim0.5$deg)
is largely free of sub-clustering, 
we see that there is a significant grouping to the North
of the main cluster overdensity (RA$\approx 198$, Dec$\approx$40.8).
We perform a search of the NASA Extra-Galactic Database (NED)
to attempt to identify this.  The best match is cluster
entry number 632 from Yoon et al.\ (2008; redshift $z=0.0677$, 
at $0.95\sigma_z$ from the mean redshift of Abell 1691).
This group is therefore probably a minor infalling group to Abell~1691
that has been picked up by the DS algorithm.

Also apparent from Fig.~\ref{fig:subs} and Fig.~\ref{fig:structure}
are several (arguably 3 or 4) 
filaments that link to the core of Abell~1691 -- particularly 
the eastern extension of the cluster core.  Given the
velocity dispersion of the cluster, it is not unexpected
to see this many filaments connecting to it (Pimbblet et al.\ 2004).

\section{AGN}
We now turn to the active galaxy population.
To determine which cluster members are AGN, we make use
of a BPT diagram (Baldwin, Phillips \& Terlevich 1981).  
By plotting flux ratio [OIII]/H$\beta$ against [NII]/H$\alpha$,
Baldwin et al.\ (1981) are able to
effectively separate AGN from star-forming galaxies.
Implicitly, the use of the BPT diagram means that we are
identifying type 2 AGN (those AGN that are obscured by a
dusty circumstellar medium and give rise to strong
narrow emission lines) rather than type 1 AGN (a direct 
view of the broad line region).  We check this is
true by confirming that all galaxies classed as AGN by
the BPT diagram have an SDSS spectroscopic classification
of `galaxy' (as opposed to `qso').

\begin{figure}
\centerline{\psfig{file=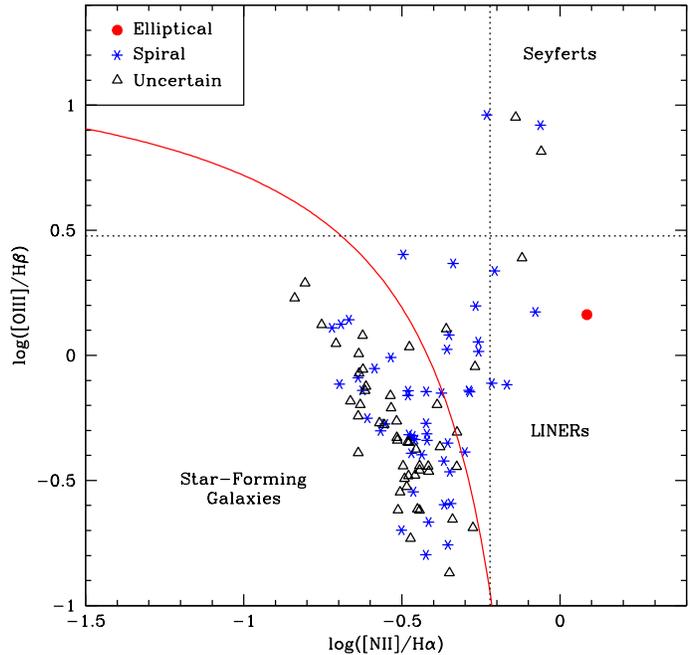,angle=0,width=3.75in}}   
\caption{BPT diagram for all cluster members with S/N$>3$ in each
line.  The Kauffmann et al.\ (2003) division between AGN and
star-forming galaxies is the solid curve, and we also mark
the loci of Seyferts ([OIII]/H$\beta > 3$ and [NII]/H$\alpha>0.6$)
and LINERs ([OIII]/H$\beta < 3$ and [NII]/H$\alpha>0.6$) with the
vertical and horizonal dotted lines.  The point types correspond
to the Galaxy Zoo morphological classifications of these galaxies.
}
        \label{fig:bpt}
\end{figure}

In Fig.~\ref{fig:bpt} we show the BPT diagram for
our cluster members that have S/N$>3$ in
the requisite lines.  To split the AGN from 
star-forming galaxies, we use the Kauffmann et al.\ (2003; 
see also 
Stasi{\'n}ska et al.\ 2006;
Brinchmann et al.\ 2004;
Kewley et al.\ 2001; Veilleux \& Osterbrock 1987)
relation for AGN: 
$log_{10}([OIII]/H\beta)>0.61 / \{log_{10}([NII]/H\alpha) -0.05\}+1.3$.
Although we could extend our analysis to weaker AGN classes
and `retired' AGN (Cid Fernandes et al.\ 2010), we elect to
stick with the Kauffmann et al.\ approach due to its wider adoption
in the literature.
We also draw on the definitions of Seyferts and LINERs (see Kauffmann et al.\ 2003;
Ho, Filippenko \& Sargent 1997).
Our BPT diagram is typical of the general galaxy population, including
a LINER galaxy with an elliptical morphology (cf.\ Heckman 1980).
A visual check of the SDSS 
imaging confirms the Galaxy Zoo morphology class:
this galaxy is indeed an early-type (possibly S0).

Is the number of AGN that we find in Abell 1691 typical?
To address this question, we note that
Popesso \& Biviano (2006) make a study of several hundred 
galaxy clusters and find that the fraction of AGN anti-correlates with
$\sigma_z$. Explicitly, they report that 
$f_{AGN} = (-1.21 \pm 0.12) log(\sigma_z) + (2.50 \pm 0.24)$
(Popesso \& Biviano 2006).
This equation predicts an AGN fraction of 0.07 for A1691
using $\sigma_z = 1009$ kms$^{-1}$, and 0.22 using
$\sigma_z = 402$ kms$^{-1}$ (see above for the derivation of these 
two values of $\sigma_z$).
We note their definition of the fraction
of AGN contained in a cluster is the number of AGN divided by 
the total galaxy population within $r_{200}$ and brighter than
$M_r = -20.0$.
We compute $r_{200}$ from Girardi et al.\ (1998):
$r_{200} \sim r_{virial} = 0.002 \sigma_z = 2.01$ Mpc.  At 
the mean redshift of Abell 1691, this radius corresponds to
25 arcmin.  
For convenience (and to be in-line with the calculation of $f_B$, below)
we stick to our magnitude limit of $r=17.77$.
We obtain $f_{AGN} = 5/75 = 0.06\pm0.03$ (where the
quoted error is simply a Poissonian error).  This fraction
is consistent with the (anti-correlation) equation reported by 
Popesso \& Biviano (2006; above) that predicts $f_{AGN}=0.07$ for 
our cluster if we adopt $\sigma_z = 1009$ kms$^{-1}$, and is more than
$3\sigma$ away from the AGN fraction predicted for
$\sigma_z = 402$ kms$^{-1}$.
Therefore this firms our above argument that $\sigma_z = 1009$ kms$^{-1}$
is the correct velocity dispersion for the cluster.

\section{Colour-magnitude relation}
\begin{figure}
\centerline{\psfig{file=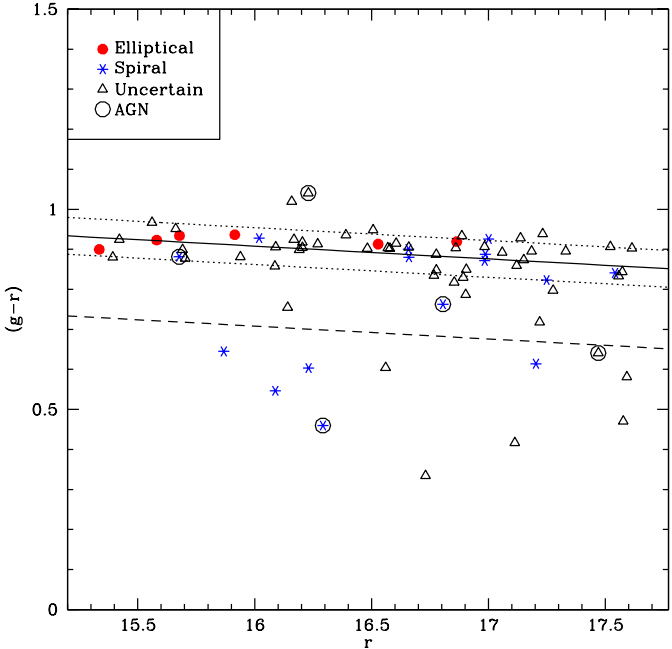,angle=0,width=3.75in}}
\caption{Colour-magnitude relation for cluster members
within $r_{200}$ of the cluster centre. 
Symbols are the same as for the BPT diagram (Fig.~\ref{fig:bpt}), 
and we circle our identified AGN.
The solid line displays the fit to the
red sequence with the scatter in the fit's intercept denoted
by the parallel dotted lines.  The dashed lines denote the limits used
to define $f_B$ (i.e.\ blue galaxies are to the left and below these 
lines).
}
        \label{fig:cmr}
\end{figure}

On a colour-magnitude plane, a galaxy cluster exhibits a strong
correlation for its early-type members: the so-called red sequence
(e.g.\ Visvanathan \& Sandage 1977; Bower, Lucey \& Ellis 1992).
Fig.~\ref{fig:cmr} displays the r vs.\ (g-r) colour-magnitude
diagram for cluster members within $r_{200}$ of the cluster centre.
We fit the red sequence as per Jensen \& Pimbblet (2012), using
the Lorentzian merit function minimized using the Nelder-Mead 
down-hill simplex algorithm (Press et al. 1992).  The scatter in the
intercept of the red sequence fit is displayed as parallel lines
around the red sequence fit in Fig.~\ref{fig:cmr}, along with
the Galaxy Zoo morphology classifications and identified AGN.

We use Fig.~\ref{fig:cmr} to define the blue fraction ($f_B$) of
galaxy cluster members in an analogous way to Butcher \& Oemler (1984).
Formally, Butcher \& Oemler defined $f_B$ 
to be the fraction of galaxies that are
brighter than $M_V=-20$ and have a rest-frame colour that is 0.2 in (B-V) bluer
than the fitted colour-magnitude relation.  
However, since we are only interested in comparing relative values (rather
than comparing $f_B$ to external works), we simplify this definition
for our convenience by defining $f_B$ to be those galaxies 0.2 in (g-r)
bluer than the fitted red sequence as our definition of blue.
For the magnitude limit, we convert $M_V=-20$ to $r$-band
using one of the stellar magnitude conversions presented in Jester et al.\ (2005):
$V = g - 0.59\times(g-r) - 0.01$ and assuming a `typical' colour for our galaxies
(namely the red sequence colour at $M^{\star}$).  This gives a limit
of $r\approx18.3$. This is fainter than our r-band selection limit for our
sample and we therefore use all galaxies in our sample for $f_b$ calculations, below.
Additionally, we note that this is an invalid conversion since
we have not used point sources and made assumptions about the typical galaxy
colour. We re-emphasize that we are only interested in the relative trend
within our own dataset rather than an absolute comparison to external datasets
and therefore our approach is sound for this purpose.

\begin{figure}
\vspace*{-3cm}
\centerline{\psfig{file=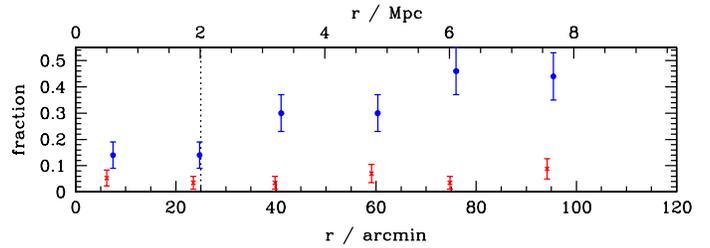,angle=0,width=3.75in}}
\vspace*{-3cm}
\caption{Radial variation of the blue fraction, $f_B$ (blue circles),
and the AGN fraction, $f_{AGN}$ (red crosses), in Abell 1691.  
The vertical dotted line denotes $r_{virial}$ and each
point contains the same number of galaxies (with $f_B$ arbitrarily 
offset slightly from $f_{AGN}$ for the sake of clarity).  The 
two different fractions follow the same (weak) trend to an approximation --
the cluster core being hostile to both blue galaxies and AGN.
}
        \label{fig:fab}
\end{figure}

In Fig.~\ref{fig:fab} we plot the variation of $f_B$, by our
definition, with radius from the cluster centre.  We complement this
trend with the radial variation of $f_{AGN}$ (as defined in Section~4).
The trend in $f_B$ is to increase away from the cluster core,
but for $f_{AGN}$, it looks reasonably flat until a very high radius from
the cluster centre; and even there the $f_{AGN}$ trend is not significant.
The best statement we can make is that 
the cluster core appears hostile to both bright blue galaxies
and active galaxy types.

Many other authors have previously noted the lack of bright blue galaxies 
in the cores of clusters and the (weak) radial trend for $f_B$ 
(Raichoor \& Andreon 2012a;
Li et al.\ 2009; Tran et al.\ 2005; Dahl{\'e}n et al.\ 2004;
Dahl{\'e}n, Fransson \& N{\"a}slund 2002; Kodama \& Bower 2001; 
see also Chung et al.\ 2011 who notes that `the 
fraction of star-forming galaxies increases with cluster radius
but remains below the field value even at $3r_{200}$')
for a given absolute magnitude limit used to define the Butcher-Oemler effect.  
Our results for Abell~1691 support these previous works and add weight to
the arguments presented elsewhere (Wilman, Zibetti,
\& Budav{\'a}ri 2010;
Wetzel, Tinker \& Conroy 2012;
Hou et al.\ 2012)
that the red galaxy fraction (i.e.\ $1-f_B$) correlates 
with environment only up to $\sim 1$ Mpc away from the cluster core. 
A tantalizing facet of Fig.~\ref{fig:fab}
is the upturn of $f_B$ at 2--3 Mpc from the cluster core.
Although statistically not significant, we note that 
this radius is about the same seen by Porter et al.\ (2008) who report
a fractional enhancement in star-bursting galaxies at these radii along
filaments of galaxies. We therefore 
hypothesize that this maxima is a reflection of this 
enhanced star-formation rate seen by Porter et al.\ (2008)
and probably caused by first-time 
harassment (Moore et al.\ 1996) of galaxies within those filaments.
A much larger sample of galaxy clusters will be required to probe this
blue fraction enhancement and definitively ascertain whether such 
an enhancement is significant.

The radial variation for $f_{AGN}$ has also been noted by
other authors (Hwang et al.\ 2012; 
Gilmour et al.\ 2007;
Johnson, Best \& Almaini 2003; Kauffmann et al.\ 2003), but there are yet
others who suggest that the AGN distribution may be flat (as we find out to 
several Mpc), or even
concentrated toward the cluster centres (Branchesi et al.\ 2007;
Martini et al.\ 2007; Ruderman \& Ebeling 2005). Our results support
a mild increase in AGN fraction at large radii away from the cluster
centre, but otherwise suggest a flat distribution.  
This result supports the notion that AGN incidence may be driven
by galaxy-galaxy interactions when there is an available gas supply
to fuel the AGNs.

\section{Spectroscopic classification}

We further quantify the galaxy population by classifying the
cluster members on the basis of the line strengths of two
key diagnostic features: $H\delta$ and [OII] (Poggianti et al.\ 1999;
Dressler et al.\ 1997; see also Pimbblet et al.\ 2006).  In combination,
these two lines can successfully discriminate between a number
of important stages in galaxy evolution, including the E+A phase
(i.e.\ strong $H\delta$ absorption coupled with a lack of current
star formation as indicated by a lack of [OII] emission).  Although
we could have used $H\alpha$ as our indicator of present star-formation,
we elect to employ this classification scheme to be in-line with previous
studies.  A summary of the classification scheme along with
its broad interpretations is presented in
Table~\ref{tab:class}.

\begin{figure*}
\centerline{\psfig{file=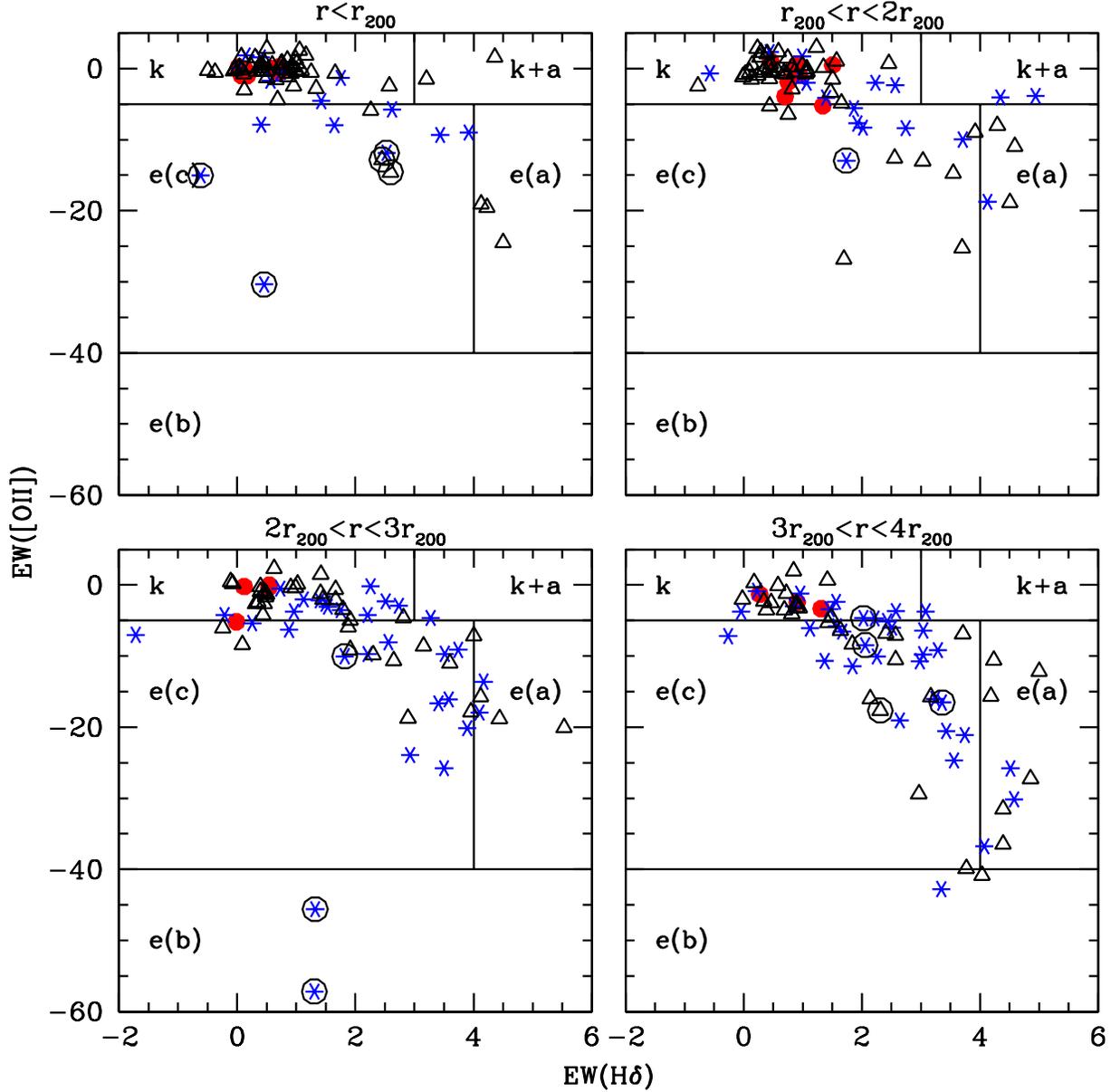,angle=0,width=6.75in}}
\vspace*{-0.5cm}
\caption{Spectroscopic classifications of the cluster
galaxy population according
to the Poggianti et al.\ (1999) scheme which is based on
EW measurements of $H\delta$ and [OII]
and which we split by radius from the cluster centre.  
By convention, emission
lines are reported with negative values.
Morphological types from Galaxy Zoo
are marked with different symbols and AGN are circled (as per
Fig.~\ref{fig:cmr}).
}
        \label{fig:classes}
\end{figure*}

\begin{table*}
\begin{center}
\caption{Spectroscopic classification scheme used in this work
and its broad interpretation (see Poggianti et al.\ 1999; see also Poggianti et al.\ 2001).
A further class of a+k is noted by Poggianti et al.\ (1999) but
we do not use it here as we have no galaxies that would be classified
as an a+k type.
\hfil}
\begin{tabular}{lll}
\hline
Class & Definition & Interpretation \\
\hline
k	& EW($H\delta$)$<3\rm{\AA}$ and EW([OII])$>-5\rm{\AA}$ & Passive: star-formation ceased
$\sim$several Gyr ago. \\
k+a 	& EW($H\delta$)$>3\rm{\AA}$ and EW([OII])$>-5\rm{\AA}$ & Post-starburst: a young population
of A-stars with no present star-formation.\\
e(c)	& EW($H\delta$)$<4\rm{\AA}$ and $-5>$EW([OII])$>-40\rm{\AA}$ & Star-forming galaxy; 
similar to a present day, steadily star-forming, spiral galaxy.\\
e(a)	& EW($H\delta$)$>4\rm{\AA}$ and $-5>$EW([OII])$>-40\rm{\AA}$ & Recent, strong star-burst
galaxy whose star-formation rate is falling back down.\\
e(b)	& EW([OII])$<-40\rm{\AA}$ & On-going star-burst.\\
\hline
\end{tabular}
  \label{tab:class}
\end{center}
\end{table*}

In Fig.~\ref{fig:classes} we display the spectroscopic classifications
of cluster members split by radial bin from the cluster core.
As would be 
expected, the passive k-type galaxies dominate at the cluster core (Table~\ref{tab:frac})
and gradually fall-off with increasing radius away from the cluster centre (indeed, the
outermost radial bin may be regarded as the `field' value).  
The values for the total emission classes versus non-emission classes are very comparable
to the values reported by Pimbblet et al.\ (2006) for massive, highly X-ray luminous
clusters (a non-emission class fraction of 0.82 within $r_{200}$ and 0.74 
beyond $r_{200}$) which are in turn a higher fraction than high redshift massive
clusters (Poggianti et al.\ 1999).

\begin{table*}
\begin{center}
\caption{Radial fractions of different spectroscopic classes.  The final two rows
give the summed emission class fractions (e(b) plus e(a) plus e(c)) and
non-emission class fractions (k plus k+a classes). All errors are Poisson.
\hfil}
\begin{tabular}{llllll}
\hline
Class & $r<r_{200}$ & $r_{200}<r<2r_{200}$ & $2r_{200}<r<3r_{200}$ & $3r_{200}<r<4r_{200}$ \\
\hline
k       & $0.80 \pm 0.10$ & $0.70 \pm 0.10$ & $0.54 \pm 0.10$ & $0.39 \pm 0.07$ \\  
k+a     & $0.03 \pm 0.02$ & $0.03 \pm 0.02$ & $0.01 \pm 0.01$ & $0.01 \pm 0.01$ \\
e(c)    & $0.14 \pm 0.04$ & $0.21 \pm 0.05$ & $0.33 \pm 0.07$ & $0.45 \pm 0.08$ \\
e(a)    & $0.04 \pm 0.02$ & $0.05 \pm 0.03$ & $0.09 \pm 0.03$ & $0.13 \pm 0.04$ \\
e(b)	& nil & nil & $0.03 \pm 0.02$ & $0.03 \pm 0.02$ \\
\hline
Emission Classes	& $0.18 \pm 0.06$ & $0.26 \pm 0.06$ & $0.45 \pm 0.08$ & $0.61 \pm 0.07$ \\
Non-emission Classes	& $0.83 \pm 0.11$ & $0.73 \pm 0.11$ & $0.55 \pm 0.10$ & $0.40 \pm 0.10$ \\
\hline
\end{tabular}
  \label{tab:frac}
\end{center}
\end{table*}

\subsection{Red and passive spirals}
Figures~\ref{fig:cmr} and~\ref{fig:classes} contain a number of spiral
galaxies that are red and (or) passive.  Since spiral galaxies possess young stellar 
populations, the existence of these red spiral galaxies is potentially,
at first glance, surprising.
However, the colour of a galaxy and its morphology are likely to be driven by
different mechanisms: the colour is a reflection of the recent star-formation
history (say, within the past $\sim$Gyr) whilst the morphology is more
sensitive to recent kinematic or dynamical events in the galaxies' life 
(cf.\ Boselli \& Gavazzi 2006).
Red and passive spiral galaxies within clusters
may therefore represent an important, but brief, transition
phase in the life of a galaxy: a point at which its star-formation rate has
been truncated (or all the star-forming gas has been used up) 
but it is yet to undergo significant morphological transformation to an
S0 or elliptical morphology
(Goto et al.\ 2003b;
Moran et al.\ 2006;
Wolf et al.\ 2009; Bundy et al.\ 2010; Lee et al.\ 2010;
Masters et al.\ 2010; see also Mahajan \& Raychaudhury 2009;  Ishigaki et al.\ 2007).

To investigate the relation of red, passive spiral galaxies with environment, 
we now extract them from our catalogue using the following criteria: (i) they
must be of the `k' spectroscopic class (Table~\ref{tab:class}); (ii) they must 
be spiral as defined by Galaxy Zoo; (iii) they must be photometrically red -- within
the dotted parallel 
lines that represent the scatter of the colour-magnitude relation's
fitted intercept displayed on Fig.~\ref{fig:cmr}.  This yields 18
red and passive spiral galaxies cluster members within $4r_{200}$.
Within $r_{200}$, 5/15 sprial galaxies are red and passive ($0.33 \pm 0.15$).
This changes to 13/81 ($0.16 \pm 0.05$) between $r_{200}$ and $4r_{200}$.
Expressed as a fraction of the total cluster population,
the red and passive spiral galaxies constitute a fraction of $\sim0.06$
at all radii.
This suggests a lack of significant
environmental dependence for the location of red and passive spiral
galaxies and agrees with Masters et al.\ (2010)
who note that although red spirals may prefer intermediate environs,
`environment alone is not sufficient to determine if a galaxy will become a red spiral'.
Gallazzi et al.\ (2009) argue that red star-forming galaxies seem ubiquitous 
at all galaxy densities. Given the relationship between radius and density for 
cluster galaxies, this supports our result of a none varying fraction 
of red passive sprirals with radius.  This means that an environmental mechanism
such as harassment may not be required to explain such galaxies: a quenching
event followed by a slow morphological transformation is likely to be sufficient 
to cause these observations (Wolf et al.\ 2009). This supports Bamford et al.\ (2009)
who argue that the timescale for morphological transform is longer
than that for colour transform on the basis of an environmental dependence with 
galaxy colour at fixed morphology from Galaxy Zoo (Lintott et al.\ 2008).

\section{Mass Relationships}
The above analysis has considered the relationship between
various parameters, such as AGN fraction, to environment -- proxied
as the distance from the cluster centre\footnote{Distance to the cluster
centre can also be considered a temporal measure -- 
i.e.\ the time since infall (Gao et al.\ 2004); but such a signal
may be contaminated (e.g.\ Pimbblet 2011).}.
To better explore the covariance of environment and galaxy mass,
we now re-do that above analysis but use galaxy stellar mass
as our primary abscissa.  We obtain the galaxy stellar mass for
our cluster members from the SDSS value-added catalogue 
(see www.mpa-garching.mpg.de/SDSS/DR7/Data/stellarmass.html).
Although the original stellar mass estimates by Kauffmann et al.\ (2003b)
are based on the EW of $H\delta$ and the D4000 Angstrom break, the data
used here is based on the fit to the SDSS photometry which, like 
Kauffmann et al.\ (2003b),
uses a large set of Bruzual \& Charlot (2003) models to derive 
its output.  
The distributors of the dataset note that there are
some differences to the line index fit derived stellar masses of 
Kauffmann et al.\ (2003b), including better accuracy at
lower masses (see: www.mpa-garching.mpg.de/SDSS/DR7/mass\_comp.html).
We therefore contend that these masses should be more than robust 
enough for our purposes of correlating AGN fraction (etc.) with mass.

\begin{figure}
\centerline{\psfig{file=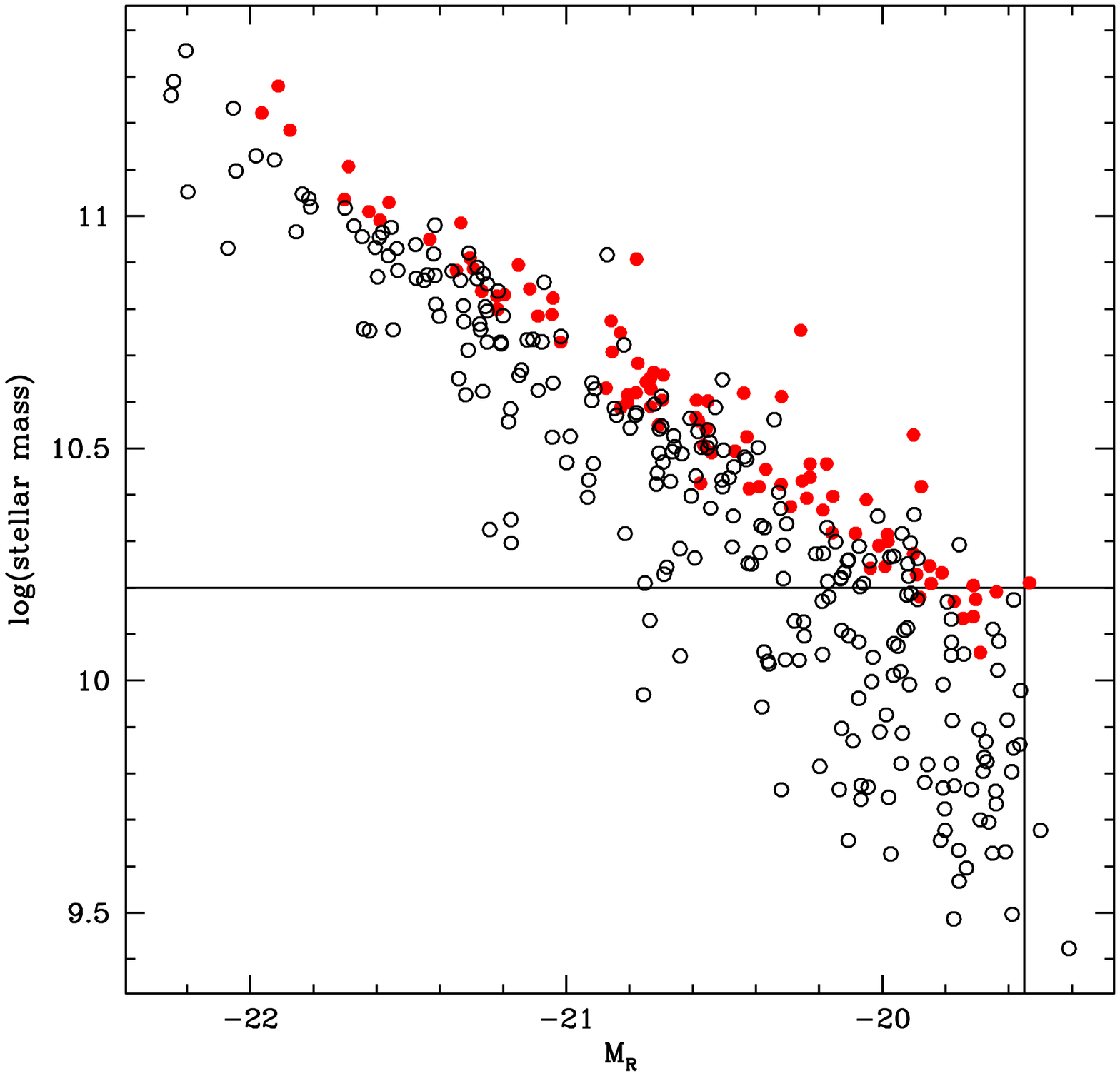,angle=0,width=3.75in}}
\caption{Galaxy mass versus absolute magnitude for cluster members.
The red filled circles are the red sequence galaxies.  The horizontal
line denotes the mass limit at which we are complete for the reddest 
galaxies (i.e.\ red sequence galaxies, or redder, 
as defined by the scatter around the line of best fit in Fig.~\ref{fig:cmr}), 
whilst the vertical line is $r=17.77$
at the mean cluster redshift.  We avoid biasing our mass relation analyses
by using only galaxies in the upper left (high mass, brighter) quadrant
of this plot.
}
        \label{fig:correction}
\end{figure}

Before we repeat the above analysis for $f_{AGN}$ and $f_B$ using galaxy
stellar mass (as above), we must guard
against the selection bias of having a top-end mass limit with a 
low-end $r$-band magnitude limit (cf.\ Holden et al.\ 2007).
We therefore construct a mass versus absolute magnitude diagram
and find the mass limit at which we are complete for the reddest galaxies
(i.e.\ colour-magnitude red sequence galaxies) and the absolute magnitude
limit corresponding to
$r=17.77$ at the mean cluster redshift (Fig.~\ref{fig:correction}).
These limits are log(stellar mass)=10.2 and $M_R=-19.55$ respectively and
we only use galaxies more massive and brighter than these limits 
(Fig.~\ref{fig:correction}) in the subsequent analysis.
This ensures that we will be complete for these galaxies, regardless of
star formation history.

\begin{figure}
\vspace*{-3cm}
\centerline{\psfig{file=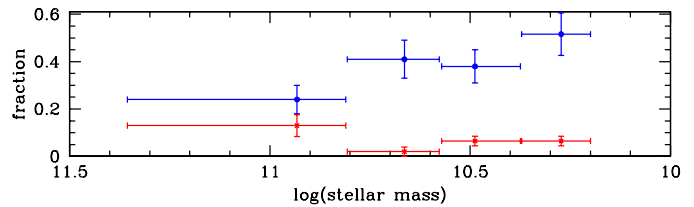,angle=0,width=3.75in}}
\vspace*{-3cm}
\caption{As for Fig.~\ref{fig:fab}, but the blue fraction (blue circles)
and AGN fraction (red crosses) are plotted as a function of galaxy stellar
mass.  The errors in the y-direction are Poissonian.  However, the `errors' in the
x-direction simply denote the full range of galaxy 
stellar mass that each point is derived from.
The blue fraction is driven by low mass galaxies ($\sim3\sigma$ level)
whereas the AGN fraction appears mostly driven by high mass galaxies 
($\sim2\sigma$ level).
}
        \label{fig:fab2}
\end{figure}

The results
are displayed in Fig.~\ref{fig:fab2}. There are two striking features
about this plot.  Firstly the blue fraction, $f_B$, appears to be
primarily driven by low stellar mass galaxies (log(stellar mass)$<10.8$).
The difference between the lowest mass bin and the highest is significant at
a $\sim3 \sigma$ level for $f_B$.
Although the trend in AGN fraction suggests that $f_{AGN}$ is being
driven by the higher mass regimes in agreement with previous works 
(Xue et al.\ 2010; Brusa et al.\ 2009) and 
unsurprising given that supermassive black holes are
almost exclusively found in massive galaxies (Magorrian et al.\ 1998), 
our limited sample is only able
to place a significance of $\sim 2 \sigma$ on this trend.  In both cases, 
these trends are much stronger than found in Fig.~\ref{fig:fab}.
The inference of this result for the Butcher-Oemler effect is 
that it is almost exclusively driven by lower mass
(log(stellar mass)$<10.8$) galaxies.  
This is supported by Tajiri 
\& Kamaya (2001) who contend that blue Butcher-Oemler galaxies 
are the less-massive cluster galaxies that also 
possess small (cluster frame) peculiar velocities and De Propris
et al.\ (2003) who note that the 
Butcher-Oemler effect is primarily due to low-luminosity galaxies 
that are presently star-forming.

\begin{table*}
\begin{center}
\caption{As for Table~\ref{tab:frac}, but for stellar mass fractions 
of the different spectroscopic classes.  
\hfil}
\begin{tabular}{llllll}
\hline
Class   & $11.5>log$(stellar mass)$>10.8$ & $10.8>log$(stellar mass)$>10.5$ & $10.5>log$(stellar mass)$>10.2$ \\
\hline
k       & $0.79 \pm 0.11$ & $0.82 \pm 0.10$ & $0.62 \pm 0.08$  \\
k+a     & nil		  & $0.04 \pm 0.02$ & $0.01 \pm 0.01$  \\
e(c)    & $0.17 \pm 0.05$ & $0.14 \pm 0.04$ & $0.33 \pm 0.06$  \\
e(a)    & nil		  & nil               & $0.04 \pm 0.02$  \\
e(b)	& $0.03 \pm 0.02$ & nil               & nil \\
\hline
Emission Classes        & $0.20 \pm 0.05$ & $0.14 \pm 0.04$ & $0.37 \pm 0.06$ \\
Non-emission Classes    & $0.79 \pm 0.11$ & $0.86 \pm 0.10$ & $0.63 \pm 0.08$ \\ 
\hline
\end{tabular}
  \label{tab:mfrac}
\end{center}
\end{table*}

In Table~\ref{tab:mfrac} we compute the spectroscopic classifications for
our galaxies (as per Table~\ref{tab:frac}), but split on galaxy mass using the 
bias corrected sample (Fig.~\ref{fig:correction}).
Immediately we see that the passive k types are dominant at the high mass
regime (log(stellar mass)$>10.5$).  This supports the notion that massive galaxies
are mostly non-star-forming and dead, regardless of their environment.  
Meanwhile, of the emission line classes, the recent star-burst galaxies (e(a) types; 
but also see Poggianti et al.\ 2001 who suggest 
the e(a) class may include a significant 
contribution from on-going dust-enshrouded star-bursts)
are exclusively found at lower mass regimes.  

We now turn to the red and passive spiral galaxies in the bias corrected sample.
Out of all spiral galaxies, we find that 6/37 ($0.16\pm0.07$) 
are red and passive at log(stellar mass)$<$10.5, whereas this grows
to 12/48 ($0.25\pm0.07$) at log(stellar mass)$>$10.5.  In terms of the whole
galaxy population, these fractions become $0.06\pm0.01$ and $0.08\pm0.02$ respectively
(where the errors are Poissonian).
Although not highly significant (i.e.\ less than $2\sigma$ difference), 
this provides the tantalizing 
suggestion that the red and passive population is driven by higher
mass rather than environment.  This agrees with observations made by Masters et al.\ (2010)
who show that red galaxies are predominantly massive galaxies, regardless of
their morphology.  
Since Galaxy Zoo is limited to simply saying that a given galaxy is a spiral, hence a
future line of investigation would be to determine the sub-class of
spiral that these red and passive spirals are. We could then 
address whether earlier spirals (i.e.\ Sa) 
classes dominate the passive spiral fraction.

\subsection{Velocity Dispersion Profiles}

\begin{figure*}
\centerline{\psfig{file=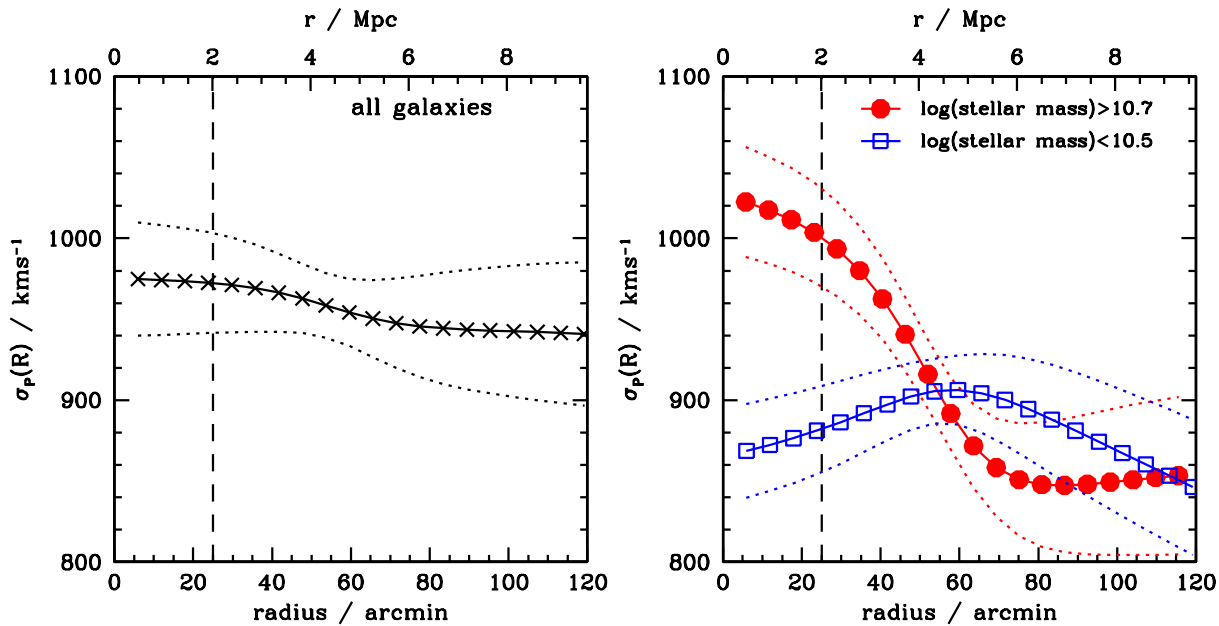,angle=0,width=6.75in}}
\vspace*{-9cm}
\caption{Velocity dispersion profile for three sets of
galaxies split by mass in A1691: high mass galaxies
(log(stellar mass)$>$10.7; filled circles), 
low mass galaxies (log(stellar mass)$<$10.5; open squares),
and all galaxies (open circles; dotted line). For clarity, we plot
these galaxies in two panels.
$r_{virial}$ is denoted by the vertical dashed line.
The error bounds (dotted lines) are $1\sigma$ 
standard deviations derived from 1000 Monte Carlo resamplings.
The high mass sample has a significantly ($>3\sigma$)
different profile to the low mass sample within $r_{virial}$.
}
        \label{fig:vdp}
\end{figure*}

Pimbblet et al.\ (2006) 
notes that the velocity dispersion
of different spectroscopic classifications of galaxies can
vary strongly -- the emission line galaxies tending to
yield larger velocity dispersions at higher radii from
the cluster centre. This ties in to work decades earlier by
(e.g.) Rood et al.\ (1972) who note different profiles for
red and blue galaxies. 
However, Pimbblet et al.\ make no comment about
how galaxy mass may affect these profiles.  Here,
we compute the velocity dispersion profile ($\sigma_P(R)$)
by mass for A1691 by following a method originally 
detailed by Bergond et al.\ (2006) for use with
globular cluster systems 
that has since found adoption in application 
to galaxy clusters (Hou et al.\ 2012; 2009).

The method involves the use of a moving window (or kernel)
that takes in to account the redshift measurements of
all cluster members for each radial bin rather than 
simply using binned radii.
The projected velocity dispersion profile is defined
as:
\begin{equation}
\sigma_P(R) = \sqrt{\frac{ \sum_i w_i(R) (x_i-\overline{x})^2 } {\sum_i w_i(R)} }
\end{equation}
where $R$ is the radius from the centre of the cluster, 
$x_i$ are the measured redshifts (or radial velocities),
$\overline{x}$ is the mean redshift of the cluster (as determined above)
and $w_i$ is the weighting fuction such that:
\begin{equation}
w_i(R) = \frac{1}{\sigma_R} exp \left( \frac{(R-R_i)^2}{2\sigma_R^2} \right)
\end{equation}
where $\sigma_R$ is the chosen width of the window (here, set to a
constant value of 30 arcmin).

The resultant velocity dispersion profiles for 
high mass (log(stellar mass)$>$10.7),
low mass (log(stellar mass)$<$10.5),
and all cluster galaxies for our bias corrected sample
is displayed in Fig.~\ref{fig:vdp}.
To obtain uncertainties on these profiles, we take a Monte Carlo
approach and bootstrap our samples 1000 times.  The error bounds
depicted in Fig.~\ref{fig:vdp} are the $1\sigma$ standard deviations
from these 1000 Monte Carlo trials.

To interpret Fig.~\ref{fig:vdp}, we note that a rising
profile would be indicative of interacting systems (cf.\ Hou et al.\ 2009;
Menci \& Fusco-Femiano 1996) or a close neighbour (Girardi et al.\ 1996).
In the case of Abell~1691, we see that the overall profile declines
gently with radius.  This underscores our earlier finding of little significant
substructure inside our cluster.  
However, the velocity dispersion profiles of the high mass and low mass
galaxies are notably different.  The high mass galaxies demonstrate a
steep decline with radius, to almost three-quarters their central value by $3r_{virial}$.
On the other hand, the lower mass galaxies show a mild increase to a local 
maxima in $\sigma_P(R)$ at
$\sim2r_{virial}$ before slowly declining again to their central levels.  
Significantly, the inner $r_{200}$ profiles for the high- and low-mass samples
are $>3\sigma$ different.
One immediate inference is that these two different mass populations have had 
significantly different accretion histories on to the cluster's potential -- the 
higher mass ones in the centre perhaps being more recent arrivals due to their
comparatively larger central $\sigma_P(R)$.
This may point to substructure in the high mass sample in the inner regions.
Therefore, to test this, we run a DS test for the high mass galaxies 
within $r_{200}$. We find 
$P(\Delta)=0.43$, meaning that there is no significant substructure for the
massive central galaxies.  
One alternative explanation
is that the lower mass galaxies can have their profile flattened quicker
than the higher mass galaxies following accretion on to the cluster.  
This would agree with the conclusions of Brough et al.\
(2008; see also Pimbblet 2008) 
who note that the timescale for new accretion on to the cluster potential is shorter
than that required to relax multiple nuclei in brightest cluster galaxies; 
more massive galaxies would seem
more able to retain a larger peculiar (to the cluster) velocity.

\subsection{Filaments, Duty Cycles and Discussion}
Filaments of galaxies have been noted in the literature as having a
pre-processing role in the evolution of galaxies that are being
funneled along them to the galaxy clusters that they inter-connect
(e.g.\ Porter et al.\ 2008; Fadda et al.\ 2008; Koyama et al.\ 2008). 
We select three of the filaments leading
off the core of A1691 by visual inspection of Fig.~\ref{fig:structure}:
one going North, in the direction of cluster candidate 632 of Yoon et 
al. (2008); the Eastward extension of A1691, and the thin curved filament
heading broadly South from the cluster.  We investigate these three filaments,
both combined and as individual objects, for trends deviating from that of the 
cluster as a whole but are unable to find an excess of 
star-burst galaxies, post-star-burst galaxies (k+a types), emission
line types, star formation rates, or specific star formation rates.
On the surface, this is contrary to the results of Porter et al.\ (2008) 
and the IR study of Fadda et al.\ (2008). 
We note however, that the 
Porter et al.\ results concern inter-cluster filaments of a straight morphology.
The only filament in our sample that could fall under that 
category in our sample is the one running to the North of the cluster core, in 
the direction of cluster candidate 632 of Yoon et 
al. (2008).  Being of the order of 5 Mpc long, this one would have been excluded from
the Porter et al.\ sample for being too short.
There is potentially an interesting inference from this null result: that filaments 
need a sufficient mass to pre-process galaxies and cause the enhanced SF observed by
Porter et al.  A larger sample of weaker (i.e.\ lower mass) filaments will 
be able to resolve this in the future (e.g., using the GAMA dataset; Driver et al.\ 2011).

\begin{figure*}
\centerline{\psfig{file=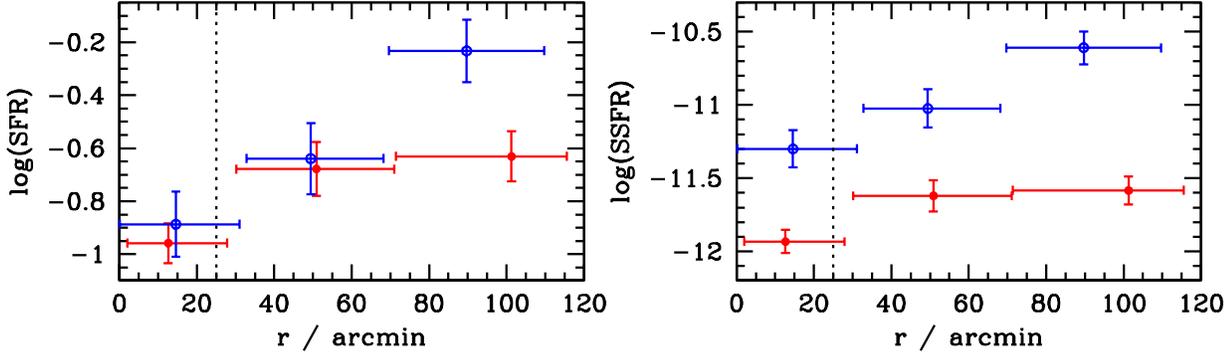,angle=0,width=6.75in}}
\vspace*{-11cm}
\caption{Star formation rate (log(SFR); left) and specific star formation rate (log(SSFR); right)
for galaxies with high stellar mass (log(stellar mass)$>10.8$; filled red circles)
and low stellar mass (log(stellar mass)$<10.5$); open blue circles) as a function of radius from
the cluster centre (the vertical dotted line denotes $r_{200}$).  Horizontal error bars denote
the range of radii sampled, the vertical errorbars are one standard deviation from a bootstrapped
sampling of the data.
The two mass regimes have significantly different forms: the higher mass 
galaxies quickly plateau in both plots with increasing radius.
}
        \label{fig:compare}
\end{figure*}

Taking the above results together, the duty cycle of lower mass galaxies appears
to be distinct from higher mass ones (Haines et al.\ 2006; von der Linden et al.\ 2010;
Oemler et al.\ 2009 and references therein).  Haines et al.\ (2006) and von der Linden
et al.\ (2010) both deduce a different environmental dependence for the star-formation
of low versus high mass galaxies that we are able to test using the data for A1691.
In Fig.~\ref{fig:compare}, we compare the star-formation rates and specific star-formation
rates of low mass (log(stellar mass)$<10.5$) and high mass (log(stellar mass$>10.7$)
galaxies as a function of radius from the cluster centre for the mass-bias corrected sample.

The two different mass regimes have significantly different forms: the high mass galaxies
plateau in both specific- and star formation rates at radii larger than $r_{200}$ whereas the
lower mass galaxies follow a strong increasing trend.  These results agree well with Haines 
et al.\ (2006) who notes that dwarf galaxies transition from being passive (i.e.\ rates consistent
with massive galaxies)
within $R_{Virial}$, to highly star forming outside this radius.

Therefore, the low mass 
galaxy population appear to undergo a starburst
event at some distance from the cluster core (more likely a large radius),
before fading in to a redder population (cf.\ Porter et al.\ 2008).  
This star-burst phase is likely to
be at least partially (perhaps mainly) responsible for the Butcher-Oemler effect
and therefore the effect may simply measure how efficiently low mass galaxies
are being processed by the cluster environment at this redshift\footnote{This 
ties in to the idea of down-sizing (Cowie et al.\ 1996).  
At higher redshifts, it may be the case that
more massive galaxies will be driving the Butcher-Oemler effect, 
but such a hypothesis requires further testing and discussion
(see, e.g., Raichoor
\& Andreon 2012a,b; De Propris et al.\ 2003), which is beyond the 
scope of this single-cluster, low redshift study.}. 
As low mass galaxies approach the
cluster core, their star-formation has likely been shut down and they will
readily be undergoing morphological transformation (although the timescale 
for morphological transform will be different to star formation truncation 
and colour change).  
Those lower mass systems in the centre of the cluster have
a much smaller velocity dispersion profile than the high mass ones (the
inverse being true at larger radii). This suggests that the low mass galaxies
at the outskirts have recently been accreted, whereas the those in the centre
have had their velocity profile dampened more quickly relative to the massive
galaxies that are better able to retain a higher peculiar velocity in the cluster
frame.
 
Meanwhile for more massive galaxies, 
star formation appears to have been shut down a long time ago in the main
part, but almost certainly
shut down prior to morphological transformation
and (potential) eventual merging with other red and dead cluster galaxies. 
Star-formation in massive galaxies (when it occurs) would therefore seem to be
driven by the gas reserves of the galaxies rather than through a star-burst phase.
We infer that the evolution of the massive galaxies is more gentle and slow 
affair compared to the brutal treatment of the low mass galaxies.

We terminate with a few caveats.  There are implicit dangers in 
plotting various parameters as a function of distance from the cluster
core due to the covariance of the 
changing morphological mixture of galaxies (Dressler 1980; see also 
Pimbblet et al.\ 2002; Andreon 2003), projection effects
(Diaferio et al.\ 2001; Rines et al.\ 2005), 
as well as the existence of
`backsplash' populations -- those galaxies
that are on radial orbits that have visited the high density
core regions of the cluster and subsequently been slung
back to larger distances away from the centre (Pimbblet 2011; 
Gill, Knebe, \& Gibson 2005).  
The latter two in particular may be problematic.  In the case of
the former, star-forming galaxies that are close to the cluster
core may simply be interlopers.  This means that the fractions
of star-forming galaxies presented in Table~\ref{tab:frac} should be
regarded as upper limits at small distances from the cluster centre. 
Secondly, the backsplash population
can cause any gradient with distance from the cluster centre to
be dampened or smeared in relation to a full three-dimensional
treatment (Atlee \& Martini 2012; Ellingson et al.\ 2001; 
Balogh, Navarro \& Morris 2000) which would render 
the points further away from the cluster core in Fig.~\ref{fig:fab} as 
lower limits.

\section{Conclusions}
We have presented an investigation of the galaxy population of Abell~1691,
an intermediate X-ray luminosity cluster from SDSS data.  Our main conclusions
can be stated as follows.\\
\ \\
\noindent $\bullet$
We have determined new global parameters for Abell~1691, including recession
velocity ($\overline{cz} =21257 \pm 54$ kms$^{-1}$) and velocity dispersion
($\sigma_{cz} = 1009^{+40}_{-36}$ kms$^{-1}$).  The cluster is morphologically
relaxed in its core regions, with no significant substructure within $r_{200}$ and
is `fed' by multiple filaments of galaxies within which a number of sub-groups reside.\\
\ \\
\noindent $\bullet$
We identify AGN from a BPT diagram and show that the cluster AGN fraction 
increases with radius from the cluster centre and is mostly driven by massive
(log(stellar mass)$>10.8$) galaxies.\\
\ \\
\noindent $\bullet$
The blue fraction of the cluster population also increases with radius away 
from the centre but is driven by lower mass galaxies 
(log(stellar mass)$<10.8$).  
Emission line galaxies follow a similar pattern. Moreover, 
the emission line galaxies are found to be driven by low mass galaxies in
the cluster (especially
the recently star-bursting e(a) class).  This suggests that some low mass galaxies
star-burst as they are accreted on to the cluster before fading to redder colours
and having their morphologies transformed.
Accordingly, we suggest that the Butcher-Oemler effect could therefore
be a consequence of mass selection effects.\\
\ \\
\noindent $\bullet$
Massive galaxies are predominantly red, occupy the centre of the cluster,
are non-star-forming,
have a different velocity dispersion profile to other cluster galaxies, 
and are likely to have had a significantly different duty
cycle compared to the bluer galaxies.  We have investigated the environment and mass of
red and passive spiral galaxies 
(which are likely to be a transition class object) and show that they are also driven 
by mass but do not have a preferential position (environment) within the cluster.\\
\ \\

This paper follows Jensen \& Pimbblet (2012) and is our second paper examining
the `environment' of intermediate $L_X$ galaxy clusters.

\section*{Acknowledgements}
We would like to thank Heath Jones, Matt Owers, Stefano Andreon, Samantha Penny 
and Tim Dolley for helpful discussion during the course of this work. KAP thanks
Iris Pimbblet for inspiration. We are also particularly grateful to the anonymous
referee for her/his feedback that significantly improved the science presented
in this work.

Figure~\ref{fig:xray} was produced using the {\sc karma} toolkit (Gooch 1996).

Funding for the SDSS and SDSS-II has been provided by the
Alfred P.\ Sloan Foundation, the Participating Institutions,
the National Science Foundation, the U.S. Department of Energy,
the National Aeronautics and Space Administration, the Japanese
Monbukagakusho, the Max Planck Society, and the Higher Education
Funding Council for England. The SDSS Web Site is http://www.sdss.org/.

The SDSS is managed by the Astrophysical Research Consortium for the
Participating Institutions. The Participating Institutions are the
American Museum of Natural History, Astrophysical Institute Potsdam,
University of Basel, University of Cambridge, Case Western Reserve
University, University of Chicago, Drexel University, Fermilab, the
Institute for Advanced Study, the Japan Participation Group, Johns
Hopkins University, the Joint Institute for Nuclear Astrophysics, the
Kavli Institute for Particle Astrophysics and Cosmology, the Korean
Scientist Group, the Chinese Academy of Sciences (LAMOST), Los Alamos
National Laboratory, the Max-Planck-Institute for Astronomy (MPIA), the
Max-Planck-Institute for Astrophysics (MPA), New Mexico State University,
Ohio State University, University of Pittsburgh, University of Portsmouth,
Princeton University, the United States Naval Observatory, and the University of Washington.

This research has made use of the NASA/IPAC Extragalactic Database (NED) which is
operated by the Jet Propulsion Laboratory, California Institute of Technology,
under contract with the National Aeronautics and Space Administration.


\begin{thebibliography}{}

\bibitem[\protect\citeauthoryear{Abazajian et 
al.}{2009}]{2009ApJS..182..543A} Abazajian K.~N., et al., 2009, ApJS, 182, 
543 

\bibitem[\protect\citeauthoryear{Abraham et 
al.}{1996}]{1996ApJ...471..694A} Abraham R.~G., et al., 1996, ApJ, 471, 694 

\bibitem[\protect\citeauthoryear{Aird et al.}{2012}]{2012ApJ...746...90A} 
Aird J., et al., 2012, ApJ, 746, 90 

\bibitem[\protect\citeauthoryear{Alexander 
\& Hickox}{2012}]{2012NewAR..56...93A} Alexander D.~M., Hickox R.~C., 2012, NewAR, 56, 93 

\bibitem[\protect\citeauthoryear{Andreon 
\& Davoust}{1997}]{1997A&A...319..747A} Andreon S., Davoust E., 1997, A\&A, 319, 747 

\bibitem[\protect\citeauthoryear{Andreon 
\& Ettori}{1999}]{1999ApJ...516..647A} Andreon S., Ettori S., 1999, ApJ, 516, 647 

\bibitem[\protect\citeauthoryear{Andreon}{2003}]{2003A&A...409...37A} Andreon S., 2003, A\&A, 409, 37 

\bibitem[\protect\citeauthoryear{Andreon et 
al.}{2006}]{2006MNRAS.365..915A} Andreon S., Quintana H., Tajer M., Galaz 
G., Surdej J., 2006, MNRAS, 365, 915 

\bibitem[\protect\citeauthoryear{Atlee et al.}{2011}]{2011ApJ...729...22A} 
Atlee D.~W., Martini P., Assef R.~J., Kelson D.~D., Mulchaey J.~S., 2011, 
ApJ, 729, 22 

\bibitem[\protect\citeauthoryear{Atlee 
\& Martini}{2012}]{2012arXiv1201.2957A} Atlee D.~W., Martini P., 2012, ApJ, submitted (arXiv:1201.2957)

\bibitem[\protect\citeauthoryear{Baldry et al.}{2006}]{2006MNRAS.373..469B} 
Baldry I.~K., Balogh M.~L., Bower R.~G., Glazebrook K., Nichol R.~C., 
Bamford S.~P., Budavari T., 2006, MNRAS, 373, 469 

\bibitem[\protect\citeauthoryear{Baldwin, Phillips, 
\& Terlevich}{1981}]{1981PASP...93....5B} Baldwin J.~A., 
Phillips M.~M., Terlevich R., 1981, PASP, 93, 5 

\bibitem[\protect\citeauthoryear{Balogh 
\& Morris}{2000}]{2000MNRAS.318..703B} Balogh M.~L., Morris S.~L., 2000, MNRAS, 318, 703 

\bibitem[\protect\citeauthoryear{Balogh, Navarro, 
\& Morris}{2000}]{2000ApJ...540..113B} Balogh M.~L., Navarro J.~F., Morris S.~L., 2000, ApJ, 540, 113 

\bibitem[\protect\citeauthoryear{Balogh et al.}{2004}]{2004ApJ...615L.101B} 
Balogh M.~L., Baldry I.~K., Nichol R., Miller C., Bower R., Glazebrook K., 
2004a, ApJ, 615, L101 

\bibitem[\protect\citeauthoryear{Balogh et al.}{2004}]{2004MNRAS.348.1355B} 
Balogh M., et al., 2004b, MNRAS, 348, 1355 

\bibitem[\protect\citeauthoryear{Bamford et 
al.}{2009}]{2009MNRAS.393.1324B} Bamford S.~P., et al., 2009, MNRAS, 393, 
1324 

\bibitem[\protect\citeauthoryear{Bergond et 
al.}{2006}]{2006A&A...448..155B} Bergond G., Zepf S.~E., Romanowsky A.~J., Sharples R.~M., Rhode K.~L., 2006, A\&A, 448, 155 

\bibitem[\protect\citeauthoryear{Bekki, Couch, 
\& Shioya}{2001}]{2001PASJ...53..395B} Bekki K., Couch W.~J., Shioya Y., 2001, PASJ, 53, 395 

\bibitem[\protect\citeauthoryear{Bekki, Couch, 
\& Shioya}{2002}]{2002ApJ...577..651B} Bekki K., Couch W.~J., Shioya Y., 2002, ApJ, 577, 651 

\bibitem[\protect\citeauthoryear{Blanton et
al.}{2003}]{2003AJ....125.2276B} Blanton M.~R., Lin H., Lupton R.~H., Maley
F.~M., Young N., Zehavi I., Loveday J., 2003, AJ, 125, 2276

\bibitem[\protect\citeauthoryear{Boselli 
\& Gavazzi}{2006}]{2006PASP..118..517B} Boselli A., Gavazzi G., 2006, PASP, 118, 517 

\bibitem[\protect\citeauthoryear{Bower, Lucey, 
\& Ellis}{1992}]{1992MNRAS.254..601B} Bower R.~G., Lucey J.~R., Ellis R.~S., 1992, MNRAS, 254, 601 

\bibitem[\protect\citeauthoryear{Branchesi et 
al.}{2007}]{2007A&A...462..449B} Branchesi M., Gioia I.~M., Fanti C., Fanti R., Cappelluti N., 2007, A\&A, 462, 449 

\bibitem[\protect\citeauthoryear{Brinchmann et 
al.}{2004}]{2004MNRAS.351.1151B} Brinchmann J., Charlot S., White S.~D.~M., 
Tremonti C., Kauffmann G., Heckman T., Brinkmann J., 2004, MNRAS, 351, 1151 

\bibitem[\protect\citeauthoryear{Brough et al.}{2008}]{2008MNRAS.385L.103B} 
Brough S., Couch W.~J., Collins C.~A., Jarrett T., Burke D.~J., Mann R.~G., 
2008, MNRAS, 385, L103 

\bibitem[\protect\citeauthoryear{Brusa et 
al.}{2009}]{2009A&A...507.1277B} Brusa M., et al., 2009, A\&A, 507, 1277 

\bibitem[\protect\citeauthoryear{Bruzual 
\& Charlot}{2003}]{2003MNRAS.344.1000B} Bruzual G., Charlot S., 2003, MNRAS, 344, 1000 

\bibitem[\protect\citeauthoryear{Bundy et al.}{2010}]{2010ApJ...719.1969B} 
Bundy K., et al., 2010, ApJ, 719, 1969 

\bibitem[\protect\citeauthoryear{Butcher 
\& Oemler}{1978}]{1978ApJ...226..559B} Butcher H., Oemler A., Jr., 1978, ApJ, 226, 559 

\bibitem[\protect\citeauthoryear{Butcher 
\& Oemler}{1984}]{1984ApJ...285..426B} Butcher H., Oemler A., Jr., 1984, ApJ, 285, 426 

\bibitem[\protect\citeauthoryear{Calvi et al.}{2012}]{2012MNRAS.419L..14C} 
Calvi R., Poggianti B.~M., Fasano G., Vulcani B., 2012, MNRAS, 419, L14 

\bibitem[\protect\citeauthoryear{Carlberg, Yee, 
\& Ellingson}{1997}]{1997ApJ...478..462C} Carlberg R.~G., Yee H.~K.~C., Ellingson E., 1997, ApJ, 478, 462 

\bibitem[\protect\citeauthoryear{Christlein 
\& Zabludoff}{2005}]{2005ApJ...621..201C} Christlein D., Zabludoff A.~I., 2005, ApJ, 621, 201 

\bibitem[\protect\citeauthoryear{Chung et al.}{2011}]{2011ApJ...743...34C} 
Chung S.~M., Eisenhardt P.~R., Gonzalez A.~H., Stanford S.~A., Brodwin M., 
Stern D., Jarrett T., 2011, ApJ, 743, 34 

\bibitem[\protect\citeauthoryear{Cid Fernandes et 
al.}{2010}]{2010MNRAS.403.1036C} Cid Fernandes R., Stasi{\'n}ska G., 
Schlickmann M.~S., Mateus A., Vale Asari N., Schoenell W., Sodr{\'e} L., 
2010, MNRAS, 403, 1036 

\bibitem[\protect\citeauthoryear{Clemens et 
al.}{2006}]{2006MNRAS.370..702C} Clemens M.~S., Bressan A., Nikolic B., 
Alexander P., Annibali F., Rampazzo R., 2006, MNRAS, 370, 702 

\bibitem[\protect\citeauthoryear{Constantin, Hoyle, 
\& Vogeley}{2008}]{2008ApJ...673..715C} Constantin A., Hoyle F., Vogeley M.~S., 2008, ApJ, 673, 715 

\bibitem[\protect\citeauthoryear{Cooper et al.}{2006}]{2006MNRAS.370..198C} 
Cooper M.~C., et al., 2006, MNRAS, 370, 198 

\bibitem[\protect\citeauthoryear{Couch 
\& Sharples}{1987}]{1987MNRAS.229..423C} Couch W.~J., Sharples R.~M., 1987, MNRAS, 229, 423 

\bibitem[\protect\citeauthoryear{Couch et al.}{1994}]{1994ApJ...430..121C} 
Couch W.~J., Ellis R.~S., Sharples R.~M., Smail I., 1994, ApJ, 430, 121 

\bibitem[\protect\citeauthoryear{Couch et al.}{1998}]{1998ApJ...497..188C} 
Couch W.~J., Barger A.~J., Smail I., Ellis R.~S., Sharples R.~M., 1998, 
ApJ, 497, 188 

\bibitem[\protect\citeauthoryear{Cowie et al.}{1996}]{1996AJ....112..839C} 
Cowie L.~L., Songaila A., Hu E.~M., Cohen J.~G., 1996, AJ, 112, 839 

\bibitem[\protect\citeauthoryear{Dahl{\'e}n et 
al.}{2004}]{2004MNRAS.350..253D} Dahl{\'e}n T., Fransson C., {\"O}stlin G., 
N{\"a}slund M., 2004, MNRAS, 350, 253 

\bibitem[\protect\citeauthoryear{Dahl{\'e}n, Fransson, {\ 
N&auml}slund}{2002}]{2002MNRAS.330..167D} Dahl{\'e}n T., Fransson C., N{\"a}slund M., 2002, MNRAS, 330, 167 

\bibitem[\protect\citename{Danese} 1980]{1980A&A....82..322D} Danese L., de
Zotti G., di Tullio G., 1980, A\&A, 82, 322

\bibitem[\protect\citeauthoryear{Deng et al.}{2010}]{2010ApJ...716..599D} 
Deng X.-F., Wen X.-Q., Xu J.-Y., Ding Y.-P., Huang T., 2010, ApJ, 716, 599 

\bibitem[\protect\citeauthoryear{De Propris et 
al.}{2003}]{2003ApJ...598...20D} De Propris R., Stanford S.~A., Eisenhardt 
P.~R., Dickinson M., 2003, ApJ, 598, 20 

\bibitem[\protect\citeauthoryear{De Propris et 
al.}{2004}]{2004MNRAS.351..125D} De Propris R., et al., 2004, MNRAS, 351, 
125 

\bibitem[\protect\citeauthoryear{Diaferio et 
al.}{2001}]{2001MNRAS.323..999D} Diaferio A., Kauffmann G., Balogh M.~L., 
White S.~D.~M., Schade D., Ellingson E., 2001, MNRAS, 323, 999 

\bibitem[\protect\citeauthoryear{di Serego Alighieri, Lanzoni, 
\& J{\o}rgensen}{2006}]{2006ApJ...647L..99D} di Serego Alighieri S., Lanzoni B., J{\o}rgensen I., 2006, ApJ, 647, L99 

\bibitem[\protect\citeauthoryear{Dressler}{1980}]{1980ApJ...236..351D} 
Dressler A., 1980, ApJ, 236, 351 

\bibitem[\protect\citeauthoryear{Dressler 
\& Gunn}{1983}]{1983ApJ...270....7D} Dressler A., Gunn J.~E., 1983, ApJ, 270, 7 

\bibitem[\protect\citeauthoryear{Dressler 
\& Shectman}{1988}]{1988AJ.....95..985D} Dressler A., Shectman S.~A., 1988, AJ, 95, 985 

\bibitem[\protect\citeauthoryear{Dressler et 
al.}{1997}]{1997ApJ...490..577D} Dressler A., et al., 1997, ApJ, 490, 577 

\bibitem[\protect\citeauthoryear{Dressler et 
al.}{1999}]{1999ApJS..122...51D} Dressler A., Smail I., Poggianti B.~M., 
Butcher H., Couch W.~J., Ellis R.~S., Oemler A., Jr., 1999, ApJS, 122, 51 

\bibitem[\protect\citeauthoryear{Dressler et 
al.}{2004}]{2004ApJ...617..867D} Dressler A., Oemler A., Jr., Poggianti 
B.~M., Smail I., Trager S., Shectman S.~A., Couch W.~J., Ellis R.~S., 2004, 
ApJ, 617, 867 

\bibitem[\protect\citeauthoryear{Dressler et 
al.}{2009}]{2009ApJ...693..140D} Dressler A., Rigby J., Oemler A., Jr., 
Fritz J., Poggianti B.~M., Rieke G., Bai L., 2009, ApJ, 693, 140 

\bibitem[\protect\citeauthoryear{Driver et al.}{2011}]{2011MNRAS.413..971D} 
Driver S.~P., et al., 2011, MNRAS, 413, 971 

\bibitem[\protect\citeauthoryear{Eastman et 
al.}{2007}]{2007ApJ...664L...9E} Eastman J., Martini P., Sivakoff G., 
Kelson D.~D., Mulchaey J.~S., Tran K.-V., 2007, ApJ, 664, L9 

\bibitem[\protect\citeauthoryear{Ebeling et
al.}{2000}]{2000MNRAS.318..333E} Ebeling H., Edge A.~C., Allen S.~W.,
Crawford C.~S., Fabian A.~C., Huchra J.~P., 2000, MNRAS, 318, 333

\bibitem[\protect\citeauthoryear{Ellingson et 
al.}{2001}]{2001ApJ...547..609E} Ellingson E., Lin H., Yee H.~K.~C., 
Carlberg R.~G., 2001, ApJ, 547, 609 

\bibitem[\protect\citeauthoryear{Fabricant, Bautz, 
\& McClintock}{1994}]{1994AJ....107....8F} Fabricant D.~G., Bautz M.~W., McClintock J.~E., 1994, AJ, 107, 8 

\bibitem[\protect\citeauthoryear{Fadda et al.}{2008}]{2008ApJ...672L...9F} 
Fadda D., Biviano A., Marleau F.~R., Storrie-Lombardi L.~J., Durret F., 
2008, ApJ, 672, L9 

\bibitem[\protect\citeauthoryear{Fisher et al.}{1998}]{1998ApJ...498..195F} 
Fisher D., Fabricant D., Franx M., van Dokkum P., 1998, ApJ, 498, 195 

\bibitem[\protect\citeauthoryear{Gallazzi et 
al.}{2009}]{2009ApJ...690.1883G} Gallazzi A., et al., 2009, ApJ, 690, 1883 

\bibitem[\protect\citeauthoryear{Gavazzi, Savorgnan, 
\& Fumagalli}{2011}]{2011A&A...534A..31G} Gavazzi G., Savorgnan G., Fumagalli M., 2011, A\&A, 534, A31 

\bibitem[\protect\citeauthoryear{Georgakakis et 
al.}{2008}]{2008MNRAS.385.2049G} Georgakakis A., et al., 2008, MNRAS, 385, 
2049 

\bibitem[\protect\citeauthoryear{Gill, Knebe, 
\& Gibson}{2005}]{2005MNRAS.356.1327G} Gill S.~P.~D., Knebe A., Gibson B.~K., 2005, MNRAS, 356, 1327 

\bibitem[\protect\citeauthoryear{Gilmour et 
al.}{2007}]{2007MNRAS.380.1467G} Gilmour R., Gray M.~E., Almaini O., Best 
P., Wolf C., Meisenheimer K., Papovich C., Bell E., 2007, MNRAS, 380, 1467 

\bibitem[\protect\citeauthoryear{Giodini et 
al.}{2012}]{2012A&A...538A.104G} Giodini S., et al., 2012, A\&A, 538, A104 

\bibitem[\protect\citeauthoryear{Girardi et al.}{1998}]{1998ApJ...505...74G} Girardi M., 
Giuricin G., Mardirossian F., Mezzetti M., Boschin W., 1998, ApJ, 505, 74

\bibitem[\protect\citeauthoryear{Girardi et 
al.}{1996}]{1996ApJ...457...61G} Girardi M., Fadda D., Giuricin G., 
Mardirossian F., Mezzetti M., Biviano A., 1996, ApJ, 457, 61 

\bibitem[\protect\citeauthoryear{G{\'o}mez et 
al.}{2003}]{2003ApJ...584..210G} G{\'o}mez P.~L., et al., 2003, ApJ, 584, 
210 

\bibitem[\protect\citeauthoryear{Gooch}{1996}]{1996ASPC..101...80G} Gooch 
R., 1996, ASPC, 101, 80

\bibitem[\protect\citeauthoryear{Goto et al.}{2003}]{2003PASJ...55..739G} 
Goto T., et al., 2003, PASJ, 55, 739 

\bibitem[\protect\citeauthoryear{Goto et al.}{2003}]{2003PASJ...55..757G} 
Goto T., et al., 2003b, PASJ, 55, 757 

\bibitem[\protect\citeauthoryear{Gr{\"u}tzbauch et 
al.}{2011}]{2011MNRAS.411..929G} Gr{\"u}tzbauch R., Conselice C.~J., Varela 
J., Bundy K., Cooper M.~C., Skibba R., Willmer C.~N.~A., 2011, MNRAS, 411, 
929 

\bibitem[\protect\citeauthoryear{Gunn 
\& Gott}{1972}]{1972ApJ...176....1G} Gunn J.~E., Gott J.~R., III, 1972, ApJ, 176, 1 

\bibitem[\protect\citeauthoryear{Haggard et 
al.}{2010}]{2010ApJ...723.1447H} Haggard D., Green P.~J., Anderson S.~F., 
Constantin A., Aldcroft T.~L., Kim D.-W., Barkhouse W.~A., 2010, ApJ, 723, 
1447 

\bibitem[\protect\citeauthoryear{Haines et al.}{2006}]{2006ApJ...647L..21H} 
Haines C.~P., La Barbera F., Mercurio A., Merluzzi P., Busarello G., 2006, 
ApJ, 647, L21 

\bibitem[\protect\citeauthoryear{Haines et al.}{2009}]{2009ApJ...704..126H} 
Haines C.~P., et al., 2009, ApJ, 704, 126 

\bibitem[\protect\citeauthoryear{Heckman}{1980}]{1980A&A....87..152H} Heckman T.~M., 
1980, A\&A, 87, 152 

\bibitem[\protect\citeauthoryear{Heckman et 
al.}{2005}]{2005ApJ...634..161H} Heckman T.~M., Ptak A., Hornschemeier A., 
Kauffmann G., 2005, ApJ, 634, 161 

\bibitem[\protect\citeauthoryear{Henriksen 
\& Byrd}{1996}]{1996ApJ...459...82H} Henriksen M., Byrd G., 1996, ApJ, 459, 82 

\bibitem[\protect\citeauthoryear{Ho, Filippenko, 
\& Sargent}{1997}]{1997ApJS..112..315H} Ho L.~C., Filippenko A.~V., Sargent W.~L.~W., 1997, ApJS, 112, 315 

\bibitem[\protect\citeauthoryear{Holden et al.}{2007}]{2007ApJ...670..190H} 
Holden B.~P., et al., 2007, ApJ, 670, 190 

\bibitem[\protect\citeauthoryear{Hou et al.}{2012}]{2012MNRAS.421.3594H} 
Hou A., et al., 2012, MNRAS, 421, 3594 

\bibitem[\protect\citeauthoryear{Hou et al.}{2009}]{2009ApJ...702.1199H} 
Hou A., Parker L.~C., Harris W.~E., Wilman D.~J., 2009, ApJ, 702, 1199 

\bibitem[\protect\citeauthoryear{Hwang et 
al.}{2012}]{2012A&A...538A..15H} Hwang H.~S., Park C., Elbaz D., Choi Y.-Y., 2012, A\&A, 538, A15 

\bibitem[\protect\citeauthoryear{Ishigaki, Goto, 
\& Matsuhara}{2007}]{2007MNRAS.382..270I} Ishigaki M., Goto T., Matsuhara H., 2007, MNRAS, 382, 270 

\bibitem[\protect\citeauthoryear{Jensen 
\& Pimbblet}{2012}]{2012MNRAS.422.2841J} Jensen P.~C., Pimbblet K.~A., 2012, MNRAS, 422, 2841 

\bibitem[\protect\citeauthoryear{Jester et al.}{2005}]{2005AJ....130..873J} 
Jester S., et al., 2005, AJ, 130, 873 

\bibitem[\protect\citeauthoryear{Johnson, Best, 
\& Almaini}{2003}]{2003MNRAS.343..924J} Johnson O., Best P.~N., Almaini O., 2003, MNRAS, 343, 924 

\bibitem[\protect\citeauthoryear{Kauffmann et 
al.}{2003}]{2003MNRAS.346.1055K} Kauffmann G., et al., 2003, MNRAS, 346, 
1055 

\bibitem[\protect\citeauthoryear{Kauffmann et 
al.}{2003}]{2003MNRAS.341...33K} Kauffmann G., et al., 2003b, MNRAS, 341, 33 

\bibitem[\protect\citeauthoryear{Kauffmann et 
al.}{2004}]{2004MNRAS.353..713K} Kauffmann G., White S.~D.~M., Heckman 
T.~M., M{\'e}nard B., Brinchmann J., Charlot S., Tremonti C., Brinkmann J., 
2004, MNRAS, 353, 713 

\bibitem[\protect\citeauthoryear{Kewley et al.}{2001}]{2001ApJ...556..121K} 
Kewley L.~J., Dopita M.~A., Sutherland R.~S., Heisler C.~A., Trevena J., 
2001, ApJ, 556, 121 

\bibitem[\protect\citeauthoryear{Kodama 
\& Bower}{2001}]{2001MNRAS.321...18K} Kodama T., Bower R.~G., 2001, MNRAS, 321, 18 

\bibitem[\protect\citeauthoryear{Koyama et al.}{2008}]{2008MNRAS.391.1758K} 
Koyama Y., et al., 2008, MNRAS, 391, 1758 

\bibitem[\protect\citeauthoryear{Lake, Katz, 
\& Moore}{1998}]{1998ApJ...495..152L} Lake G., Katz N., Moore B., 1998, ApJ, 495, 152 

\bibitem[\protect\citeauthoryear{Lavery 
\& Henry}{1988}]{1988ApJ...330..596L} Lavery R.~J., Henry J.~P., 1988, ApJ, 330, 596 

\bibitem[\protect\citeauthoryear{Larson, Tinsley, 
\& Caldwell}{1980}]{1980ApJ...237..692L} Larson R.~B., Tinsley B.~M., Caldwell C.~N., 1980, ApJ, 237, 692 

\bibitem[\protect\citeauthoryear{Lee et al.}{2010}]{2010MNRAS.403.1930L} 
Lee J.~H., Lee M.~G., Park C., Choi Y.-Y., 2010, MNRAS, 403, 1930 

\bibitem[\protect\citeauthoryear{Lewis et al.}{2002}]{2002MNRAS.334..673L} 
Lewis I., et al., 2002, MNRAS, 334, 673 

\bibitem[\protect\citeauthoryear{Li, Yee, 
\& Ellingson}{2009}]{2009ApJ...698...83L} Li I.~H., Yee H.~K.~C., Ellingson 
E., 2009, ApJ, 698, 83 

\bibitem[\protect\citeauthoryear{Lintott et 
al.}{2008}]{2008MNRAS.389.1179L} Lintott C.~J., et al., 2008, MNRAS, 389, 
1179 

\bibitem[\protect\citeauthoryear{Lintott et 
al.}{2011}]{2011MNRAS.410..166L} Lintott C., et al., 2011, MNRAS, 410, 166 

\bibitem[\protect\citeauthoryear{Lubin}{1996}]{1996AJ....112...23L} Lubin 
L.~M., 1996, AJ, 112, 23 

\bibitem[\protect\citeauthoryear{Ma 
\& Ebeling}{2011}]{2011MNRAS.410.2593M} Ma C.-J., Ebeling H., 2011, MNRAS, 410, 2593 

\bibitem[\protect\citeauthoryear{Magorrian et 
al.}{1998}]{1998AJ....115.2285M} Magorrian J., et al., 1998, AJ, 115, 2285 

\bibitem[\protect\citeauthoryear{Mahajan 
\& Raychaudhury}{2009}]{2009MNRAS.400..687M} Mahajan S., Raychaudhury S., 2009, MNRAS, 400, 687 

\bibitem[\protect\citeauthoryear{Mahajan, Haines, 
\& Raychaudhury}{2010}]{2010MNRAS.404.1745M} Mahajan S., Haines C.~P., Raychaudhury S., 2010, MNRAS, 404, 1745 

\bibitem[\protect\citeauthoryear{Margoniner et 
al.}{2001}]{2001ApJ...548L.143M} Margoniner V.~E., de Carvalho R.~R., Gal 
R.~R., Djorgovski S.~G., 2001, ApJ, 548, L143 

\bibitem[\protect\citeauthoryear{Margoniner 
\& de Carvalho}{2000}]{2000AJ....119.1562M} Margoniner V.~E., de Carvalho R.~R., 2000, AJ, 119, 1562 

\bibitem[\protect\citeauthoryear{Martini, Mulchaey, 
\& Kelson}{2007}]{2007ApJ...664..761M} Martini P., Mulchaey J.~S., Kelson D.~D., 2007, ApJ, 664, 761 

\bibitem[\protect\citeauthoryear{Masters et 
al.}{2010}]{2010MNRAS.405..783M} Masters K.~L., et al., 2010, MNRAS, 405, 
783 

\bibitem[\protect\citeauthoryear{Menci 
\& Fusco-Femiano}{1996}]{1996ApJ...472...46M} Menci N., Fusco-Femiano R., 1996, ApJ, 472, 46 

\bibitem[\protect\citeauthoryear{Miller et al.}{2003}]{2003ApJ...597..142M} 
Miller C.~J., Nichol R.~C., G{\'o}mez P.~L., Hopkins A.~M., Bernardi M., 
2003, ApJ, 597, 142 

\bibitem[\protect\citeauthoryear{Montero-Dorta et 
al.}{2009}]{2009MNRAS.392..125M} Montero-Dorta A.~D., et al., 2009, MNRAS, 
392, 125 

\bibitem[\protect\citeauthoryear{Moore et al.}{1996}]{1996Natur.379..613M} 
Moore B., Katz N., Lake G., Dressler A., Oemler A., 1996, Nature, 379, 613 

\bibitem[\protect\citeauthoryear{Moran et al.}{2006}]{2006ApJ...641L..97M} 
Moran S.~M., Ellis R.~S., Treu T., Salim S., Rich R.~M., Smith G.~P., Kneib 
J.-P., 2006, ApJ, 641, L97 

\bibitem[\protect\citeauthoryear{Nair 
\& Abraham}{2010}]{2010ApJ...714L.260N} Nair P.~B., Abraham R.~G., 2010, ApJ, 714, L260 

\bibitem[\protect\citeauthoryear{Oemler et al.}{2009}]{2009ApJ...693..152O} 
Oemler A., Jr., Dressler A., Kelson D., Rigby J., Poggianti B.~M., Fritz 
J., Morrison G., Smail I., 2009, ApJ, 693, 152 

\bibitem[\protect\citeauthoryear{Owers, Couch, 
\& Nulsen}{2009}]{2009ApJ...693..901O} Owers M.~S., Couch W.~J., Nulsen P.~E.~J., 2009, ApJ, 693, 901 

\bibitem[\protect\citeauthoryear{Pasquali et 
al.}{2009}]{2009MNRAS.394...38P} Pasquali A., van den Bosch F.~C., Mo 
H.~J., Yang X., Somerville R., 2009, MNRAS, 394, 38 

\bibitem[\protect\citeauthoryear{Pasquali et 
al.}{2010}]{2010MNRAS.407..937P} Pasquali A., Gallazzi A., Fontanot F., van 
den Bosch F.~C., De Lucia G., Mo H.~J., Yang X., 2010, MNRAS, 407, 937 

\bibitem[\protect\citeauthoryear{Patel et al.}{2011}]{2011ApJ...735...53P} 
Patel S.~G., Kelson D.~D., Holden B.~P., Franx M., Illingworth G.~D., 2011, 
ApJ, 735, 53 

\bibitem[\protect\citeauthoryear{Peng et al.}{2010}]{2010ApJ...721..193P} 
Peng Y.-j., et al., 2010, ApJ, 721, 193 

\bibitem[\protect\citeauthoryear{Pimbblet et 
al.}{2002}]{2002MNRAS.331..333P} Pimbblet K.~A., Smail I., Kodama T., Couch 
W.~J., Edge A.~C., Zabludoff A.~I., O'Hely E., 2002, MNRAS, 331, 333 

\bibitem[\protect\citeauthoryear{Pimbblet, Drinkwater, 
\& Hawkrigg}{2004}]{2004MNRAS.354L..61P} Pimbblet K.~A., Drinkwater M.~J., Hawkrigg M.~C., 2004, MNRAS, 354, L61 

\bibitem[\protect\citeauthoryear{Pimbblet et
al.}{2006}]{2006MNRAS.366..645P} Pimbblet K.~A., Smail I., Edge A.~C.,   
O'Hely E., Couch W.~J., Zabludoff A.~I., 2006, MNRAS, 366, 645

\bibitem[\protect\citeauthoryear{Pimbblet}{2008}]{2008PASA...25..176P} 
Pimbblet K.~A., 2008, PASA, 25, 176 

\bibitem[\protect\citeauthoryear{Pimbblet}{2011}]{2011MNRAS.411.2637P} 
Pimbblet K.~A., 2011, MNRAS, 411, 2637 

\bibitem[\protect\citeauthoryear{Pinkney et 
al.}{1996}]{1996ApJS..104....1P} Pinkney J., Roettiger K., Burns J.~O., 
Bird C.~M., 1996, ApJS, 104, 1 

\bibitem[\protect\citeauthoryear{Poggianti et 
al.}{1999}]{1999ApJ...518..576P} Poggianti B.~M., Smail I., Dressler A., 
Couch W.~J., Barger A.~J., Butcher H., Ellis R.~S., Oemler A., Jr., 1999, 
ApJ, 518, 576 

\bibitem[\protect\citeauthoryear{Poggianti, Bressan, 
\& Franceschini}{2001}]{2001ApJ...550..195P} Poggianti B.~M., Bressan A., Franceschini A., 2001, ApJ, 550, 195 

\bibitem[\protect\citeauthoryear{Popesso et 
al.}{2005}]{2005A&A...433..431P} Popesso P., 
Biviano A., B{\"o}hringer H., Romaniello M., Voges W., 2005, A\&A, 433, 431 

\bibitem[\protect\citeauthoryear{Popesso 
\& Biviano}{2006}]{2006A&A...460L..23P} Popesso P., Biviano A., 2006, A\&A, 460, L23 

\bibitem[\protect\citeauthoryear{Porter et al.}{2008}]{2008MNRAS.388.1152P} 
Porter S.~C., Raychaudhury S., Pimbblet K.~A., Drinkwater M.~J., 2008, 
MNRAS, 388, 1152 

\bibitem[\protect\citeauthoryear{Press et al.}{1992}]{1992nrfa.book.....P}   
Press W.~H., Teukolsky S.~A., Vetterling W.~T., Flannery B.~P., 1992,
Numerical Recipies, Cambridge University Press, Cambridge

\bibitem[\protect\citeauthoryear{Quilis, Moore, 
\& Bower}{2000}]{2000Sci...288.1617Q} Quilis V., Moore B., Bower R., 2000, Sci, 288, 1617 

\bibitem[\protect\citeauthoryear{Raichoor 
\& Andreon}{2012}]{2012A&A...537A..88R} Raichoor A., Andreon S., 2012a, A\&A, 537, A88 

\bibitem[\protect\citeauthoryear{Raichoor 
\& Andreon}{2012}]{2012A&A...543A..19R} Raichoor A., Andreon S., 2012b, A\&A, 543, A19 

\bibitem[\protect\citeauthoryear{Rakos 
\& Schombert}{1995}]{1995ApJ...439...47R} Rakos K.~D., Schombert J.~M., 1995, ApJ, 439, 47 

\bibitem[\protect\citeauthoryear{Rines et al.}{2005}]{2005AJ....130.1482R} 
Rines K., Geller M.~J., Kurtz M.~J., Diaferio A., 2005, AJ, 130, 1482 

\bibitem[\protect\citeauthoryear{Rood et al.}{1972}]{1972ApJ...175..627R} 
Rood H.~J., Page T.~L., Kintner E.~C., King I.~R., 1972, ApJ, 175, 627 

\bibitem[\protect\citeauthoryear{Ruderman 
\& Ebeling}{2005}]{2005ApJ...623L..81R} Ruderman J.~T., Ebeling H., 2005, ApJ, 623, L81 

\bibitem[\protect\citeauthoryear{Sato 
\& Martin}{2006}]{2006ApJ...647..946S} Sato T., Martin C.~L., 2006, ApJ, 647, 946 

\bibitem[\protect\citeauthoryear{Sivakoff et 
al.}{2008}]{2008ApJ...682..803S} Sivakoff G.~R., Martini P., Zabludoff 
A.~I., Kelson D.~D., Mulchaey J.~S., 2008, ApJ, 682, 803 

\bibitem[\protect\citeauthoryear{Skibba et al.}{2009}]{2009MNRAS.399..966S} 
Skibba R.~A., et al., 2009, MNRAS, 399, 966 

\bibitem[\protect\citeauthoryear{Smail et al.}{1998}]{1998MNRAS.293..124S} 
Smail I., Edge A.~C., Ellis R.~S., Blandford R.~D., 1998, MNRAS, 293, 124 

\bibitem[\protect\citeauthoryear{Spergel et
al.}{2007}]{2007ApJS..170..377S} Spergel D.~N., et al., 2007, ApJS, 170,
377

\bibitem[\protect\citeauthoryear{Stasi{\'n}ska et 
al.}{2006}]{2006MNRAS.371..972S} Stasi{\'n}ska G., Cid Fernandes R., Mateus 
A., Sodr{\'e} L., Asari N.~V., 2006, MNRAS, 371, 972 

\bibitem[\protect\citeauthoryear{Strauss et 
al.}{2002}]{2002AJ....124.1810S} Strauss M.~A., et al., 2002, AJ, 124, 1810 

\bibitem[\protect\citeauthoryear{Tajiri 
\& Kamaya}{2001}]{2001ApJ...562L.125T} Tajiri Y.~Y., Kamaya H., 2001, ApJ, 562, L125 

\bibitem[\protect\citeauthoryear{Thomas et al.}{2010}]{2010MNRAS.404.1775T} 
Thomas D., Maraston C., Schawinski K., Sarzi M., Silk J., 2010, MNRAS, 404, 
1775 

\bibitem[\protect\citeauthoryear{Tran et al.}{2003}]{2003ApJ...599..865T} 
Tran K.-V.~H., Franx M., Illingworth G., Kelson D.~D., van Dokkum P., 2003, 
ApJ, 599, 865 

\bibitem[\protect\citeauthoryear{Tran et al.}{2005}]{2005ApJ...619..134T} 
Tran K.-V.~H., van Dokkum P., Illingworth G.~D., Kelson D., Gonzalez A., 
Franx M., 2005, ApJ, 619, 134 

\bibitem[\protect\citeauthoryear{Tremonti et 
al.}{2004}]{2004ApJ...613..898T} Tremonti C.~A., et al., 2004, ApJ, 613, 
898 

\bibitem[\protect\citeauthoryear{Urquhart et 
al.}{2010}]{2010MNRAS.406..368U} Urquhart S.~A., Willis J.~P., Hoekstra H., 
Pierre M., 2010, MNRAS, 406, 368 

\bibitem[\protect\citeauthoryear{van den Bosch et 
al.}{2008}]{2008arXiv0805.0002V} van den Bosch F.~C., Pasquali A., Yang X., 
Mo H.~J., Weinmann S., McIntosh D.~H., Aquino D., 2008, MNRAS, submitted,  
(arXiv:0805.0002)

\bibitem[\protect\citeauthoryear{van der Wel et 
al.}{2007}]{2007ApJ...670..206V} van der Wel A., et al., 2007, ApJ, 670, 
206 

\bibitem[\protect\citeauthoryear{van Dokkum et 
al.}{1999}]{1999ApJ...520L..95V} van Dokkum P.~G., Franx M., Fabricant D., 
Kelson D.~D., Illingworth G.~D., 1999, ApJ, 520, L95 

\bibitem[\protect\citeauthoryear{Veilleux 
\& Osterbrock}{1987}]{1987ApJS...63..295V} Veilleux S., Osterbrock D.~E., 1987, ApJS, 63, 295 

\bibitem[\protect\citeauthoryear{Visvanathan 
\& Sandage}{1977}]{1977ApJ...216..214V} Visvanathan N., Sandage A., 1977, ApJ, 216, 214 

\bibitem[\protect\citeauthoryear{von der Linden et 
al.}{2010}]{2010MNRAS.404.1231V} von der Linden A., Wild V., Kauffmann G., 
White S.~D.~M., Weinmann S., 2010, MNRAS, 404, 1231 

\bibitem[\protect\citeauthoryear{Vulcani et 
al.}{2011}]{2011arXiv1111.0830V} Vulcani B., et al., 2012, A\&A, submitted
(arXiv:1111.0830)

\bibitem[\protect\citeauthoryear{Wetzel, Tinker, 
\& Conroy}{2012}]{2012MNRAS.424..232W} Wetzel A.~R., Tinker J.~L., Conroy C., 2012, MNRAS, 424, 232 

\bibitem[\protect\citeauthoryear{Wilman, Zibetti, 
\& Budav{\'a}ri}{2010}]{2010MNRAS.406.1701W} Wilman D.~J., Zibetti S., Budav{\'a}ri T., 2010, MNRAS, 406, 1701 

\bibitem[\protect\citeauthoryear{Wilman 
\& Erwin}{2012}]{2012ApJ...746..160W} Wilman D.~J., Erwin P., 2012, ApJ, 746, 160 

\bibitem[\protect\citeauthoryear{Wolf et al.}{2009}]{2009MNRAS.393.1302W} 
Wolf C., et al., 2009, MNRAS, 393, 1302 

\bibitem[\protect\citeauthoryear{Xue et al.}{2010}]{2010ApJ...720..368X} 
Xue Y.~Q., et al., 2010, ApJ, 720, 368 

\bibitem[\protect\citeauthoryear{Yahil
\& Vidal}{1977}]{1977ApJ...214..347Y} Yahil A., Vidal N.~V., 1977, ApJ, 214, 347

\bibitem[\protect\citeauthoryear{Yoon et al.}{2008}]{2008ApJS..176..414Y} 
Yoon J.~H., Schawinski K., Sheen Y.-K., Ree C.~H., Yi S.~K., 2008, ApJS, 
176, 414 

\bibitem[\protect\citename{Zabludoff} 1990]{1990ApJS...74....1Z} Zabludoff
A.~I., Huchra J.~P., Geller M.~J., 1990, ApJS, 74, 1

\bibitem[\protect\citeauthoryear{Zabludoff et 
al.}{1996}]{1996ApJ...466..104Z} Zabludoff A.~I., Zaritsky D., Lin H., 
Tucker D., Hashimoto Y., Shectman S.~A., Oemler A., Kirshner R.~P., 1996, 
ApJ, 466, 104 

\end{thebibliography}
\end{document}